\documentclass[journal]{IEEEtran}
\usepackage{amsmath,amssymb,amsfonts}
\usepackage{algorithmic}
\usepackage{algorithm}
\usepackage{array}
\usepackage{siunitx}
\usepackage[caption=false,font=normalsize,labelfont=sf,textfont=sf]{subfig}
\usepackage{textcomp}
\usepackage{stfloats}
\usepackage{url}
\usepackage{verbatim}
\usepackage{booktabs}
\usepackage{graphicx}
\usepackage{xcolor}

\usepackage{eurosym}
\usepackage{subcaption}
\usepackage{tikz}
\usepackage{orcidlink}
\usetikzlibrary{fit}
\usetikzlibrary{positioning, arrows.meta, shapes.multipart, calc}
\usetikzlibrary{shapes.geometric}
\usepackage{mathtools}

\usepackage[noadjust]{cite}

\newif\ifshowrevisions
\showrevisionsfalse
\definecolor{revisioncolor}{RGB}{0,0,180}
\newcommand{\revision}[1]{%
  \ifshowrevisions
    \textcolor{revisioncolor}{#1}%
  \else
    #1%
  \fi
}
\begin{document}

\title{Dynamic Flexibility Requests in Local Flexibility Markets: Quantifying the DSO Willingness to Pay}

\author{
Savvas~Panagi\,\orcidlink{0009-0008-6057-3392},~\textit{Graduate Student Member, IEEE},
Chrysovalantis~Spanias\,\orcidlink{0000-0003-3046-3287},~\textit{Member, IEEE},
and Petros~Aristidou\,\orcidlink{0000-0003-4429-0225},~\textit{Senior Member, IEEE}
\thanks{S. Panagi and C. Spanias are with the Distribution System Operator, Electricity Authority of Cyprus, Nicosia, Cyprus.}
\thanks{S. Panagi and P. Aristidou are with the Department of Electrical \& Computer Engineering \& Informatics, Cyprus University of Technology, Limassol, Cyprus.}
%
}


\maketitle

\begin{abstract}
Local Flexibility Markets (LFMs) require Distribution System Operators (DSOs) to determine both the quantity of flexibility to procure and the corresponding willingness to pay during market clearing. Existing approaches typically rely on centralized AC-OPF clearing algorithms with limited practical scalability or strictly localized, static flexibility requests driven primarily by congestion management, while the economic value of flexibility is largely neglected. This paper proposes a dynamic flexibility-request methodology in which the DSO's willingness-to-pay is embedded directly into the market-clearing objective by monetizing transformer and cable aging, network losses, and voltage congestion. To enable computationally efficient clearing, exact convex piecewise-linear epigraph reformulations of the IEEE C57.91 transformer aging model and an Arrhenius-based cable aging model are developed and formally proven to preserve exactness. The proposed framework is validated on a modified CIGRE MV benchmark and compared with a recent state-of-the-art flexibility-request methodology. The results demonstrate significantly higher market liquidity, more efficient flexibility procurement, and improved network operation while preserving transparency and non-discrimination principles of the market.

\end{abstract}

\begin{IEEEkeywords}
Local flexibility markets, distribution system operator, willingness-to-pay, transformer aging, cable aging.
\end{IEEEkeywords}

\section{Introduction}

\subsection{Motivation}

Increasing electrification and distributed resource deployment are accelerating stress on distribution networks (DNs). This rapid, frequently unpredictable expansion can outstrip the utility’s capacity to manage it, as it may occur outside the utility’s direct control and exceed its operational capabilities \cite{7155600, GONZALEZVENEGAS2021111060, 11305421}. Therefore, over the last decade, significant efforts have focused on engaging end-users through price-based incentives. However, such incentives alone have proved insufficient at higher penetration levels, as synchronized customer responses can erode demand diversity and create coincident peaks~\cite{ISGAN2025FlexPlanning}. Local Flexibility Markets (LFMs) have therefore emerged as a complementary mechanism to activate Distributed Energy Resources (DERs) while explicitly respecting local network limits \cite{LFM_EU2026}, extending wholesale market principles to the distribution level.

\subsection{Literature Review and Research Gap}

LFMs can be designed in various ways depending on regulatory frameworks and market structures \cite{jin2020local,villar2018flexibility,11690574,olivella2018optimization}. \revision{Two key design choices concern the market structure, whether the Market Operator (MO) and the Distribution System Operator (DSO) roles are unified or assigned to separate entities, and how network constraints enter market clearing~\cite{rebenaque2023success}.} The most common approach embeds network information into the clearing problem \cite{torbaghan2018market,papayiannis2025enhancement}. While computationally attractive DC approximations are widely used at the transmission level \cite{den2014dc}, they are unsuitable for DNs due to higher R/X ratios and stronger voltage sensitivities \cite{kotsonias2025operational}. As a result, network-integrated clearing often relies on AC Optimal Power Flow (AC-OPF) models or their approximations, which raise challenges related to exactness, convexity, transparency, and non-discrimination, which conflict with EU market principles \cite{farivar2013branch,li2012exact,christakou2017ac,zhou2020note,EU2019_943,EU2019_944}.

Alternatively, privacy-preserving architectures keep network assessment internal to the DSO: \revision{the DSO retains the detailed network model for its own analyses}, while the MO clears the market using only reduced inputs (e.g., flexibility requests, valuation signals, and sensitivities) \cite{fonteijn2020demonstrating,prat-network-aware}. This improves \revision{data privacy,} practicality, scalability, and transparency; the corresponding market structures are formalized in Section~\ref{sec:background}. This work adopts a request–offer architecture \revision{under the same principle, network analyses remain with the DSO, and only reduced-valuation inputs are shared with the MO}. Under this architecture, a central open question is how to quantify network-constrained flexibility requirements \cite{prat-network-aware}. The resulting decomposition between network operation and market clearing may yield overly conservative or insufficient procurement, with adverse cost and security implications, due to locational dependencies. Most studies focus on market design and clearing algorithms \cite{jin2020local,villar2018flexibility}, while flexibility needs are often treated as exogenous inputs. 

Among the first methods to explicitly quantify network-driven flexibility requirements, \cite{prat-network-aware} proposes a chance-constrained active-power OPF based on a LinDistFlow approximation to estimate bus-specific flexibility needs. Although the approach provides valuable locational signals, bus-specific procurement may restrict the number of eligible providers and thus market liquidity. To alleviate this limitation, flexibility zones are introduced to enlarge the pool of providers; however, such aggregation increases exposure to locational uncertainty and the risk of ineffective activation. These trade-offs are assessed quantitatively through the case-study comparison in Section~\ref{sec:Results_Discussions}. In a complementary direction, \cite{fonteijn2020demonstrating} aggregates flexibility quantification at the secondary-transformer level and monetizes transformer overloading through aging and loss-of-load costs. However, voltage and line-loading constraints, often the most critical operational issues in DNs, are neglected.

Related request--offer frameworks exhibit a similar incompleteness in the network representation of flexibility needs. In the transactive setting of \cite{khorasany2020framework}, flexibility is activated when community import exceeds a predefined threshold, and a distributed clearing mechanism matches the requested volume. Although the workflow resembles LFM clearing, the requirement is driven by an aggregate import limit rather than detailed distribution-network constraints. A comparable limitation appears in the multi-horizon LFM of \cite{bouloumpasis2022local}, where flexibility needs are again not derived from a comprehensive network-constrained valuation.

Finally, none of the aforementioned methodologies addresses the fundamental question: \textit{What is the DSO's willingness to pay for flexibility?} Under a request–offer–clearing market structure, the DSO must determine both the value of the requested flexibility and the corresponding bid price. A holistic quantification should therefore incorporate the impact of flexibility on all relevant network assets and operating conditions within the market-clearing process, including transformer and cable aging, network losses, and congestion management. To the best of the authors' knowledge, no existing clearing framework simultaneously considers all of these factors when determining the DSO willingness to pay.


\subsection{Paper Contributions}

This work addresses limited market liquidity, excessive localization of flexibility zones, and the absence of comprehensive monetization of multiple key network components in existing approaches. We propose a dynamic flexibility-request methodology that determines the DSO willingness to pay endogenously. The main contributions are as follows:

\begin{itemize}
    \item A \textit{dynamic flexibility request} methodology for LFMs, where the DSO willingness-to-pay is determined within the market-clearing process through the monetization of network operation, replacing static flexibility requests.
    \item An exact convex epigraph-based piecewise-linear reformulation of the IEEE C57.91 transformer and Arrhenius-based cable aging models, enabling the explicit incorporation of asset degradation into a computationally efficient market-clearing framework.
    \item A unified DSO valuation framework that jointly quantifies transformer and cable aging, medium-low voltage (MV/LV) losses, and voltage-limit violations into locational flexibility prices, validated on a modified CIGRE MV network and benchmarked against a recent state-of-the-art methodology, \revision{along with an openly released case-study dataset to support reproducible comparison}.
\end{itemize}

The remainder of the paper is organized as follows. Section~\ref{sec:background} reviews privacy-preserving and network-integrated LFM architectures. Section~\ref{sec:market_transformer_formulation} develops the convex piecewise-linear reformulations of the transformer and cable aging models. Section~\ref{sec:DSO_Network_benefits} presents the DSO willingness-to-pay valuation framework and the complete market-clearing formulation. Section~\ref{sec:case_study} describes the case-study setup, and Section~\ref{sec:Results_Discussions} discusses the numerical results and comparison with the benchmark methodology. Finally, Section~\ref{sec:conclusion} concludes the paper and outlines future research directions.

\section{Background}
\label{sec:background}

\begin{figure}
    \centering
    \includegraphics[width=1\linewidth]{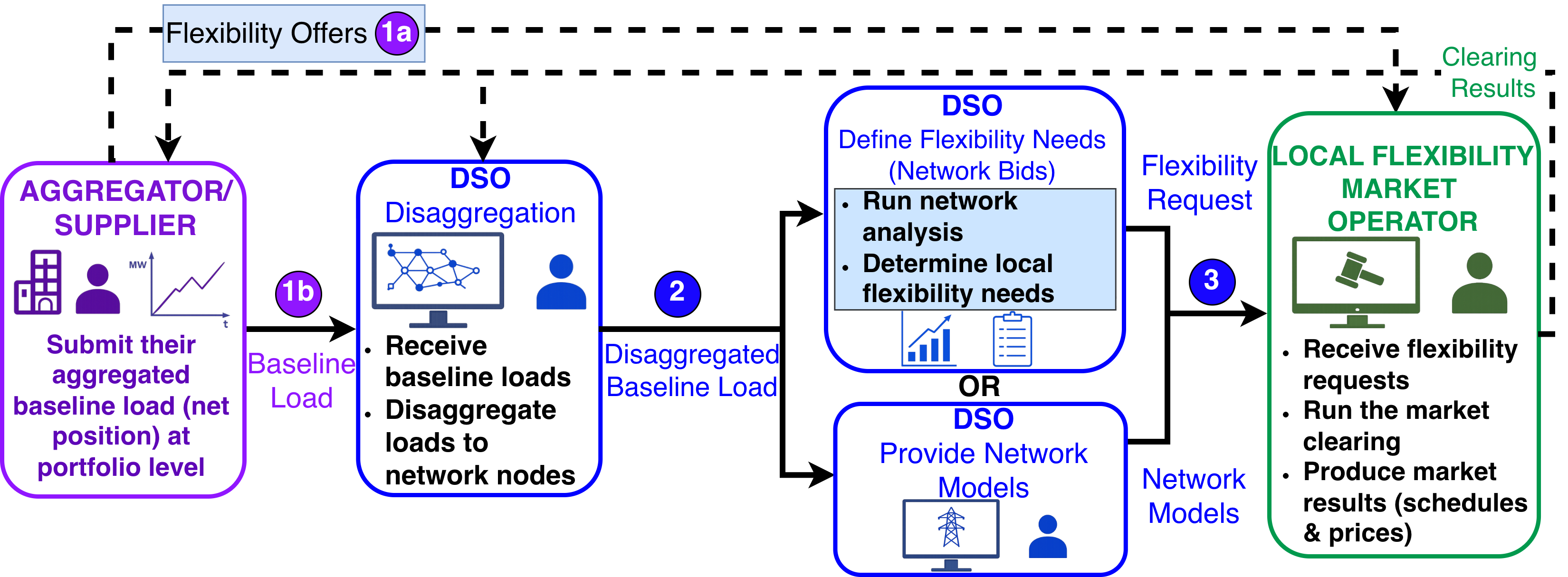}
    \caption{General workflow of a LFM, illustrating the interactions among aggregators, the DSO, and the market operator.}
    \label{fig:lf_market_designs}
\end{figure}

In a typical LFM, the DSO identifies flexibility requirements, Flexibility Service Providers (FSPs) submit offers (volume, direction, location, and price), and the MO clears the market according to the adopted objective and operational requirements \cite{jin2020local}. As illustrated in Fig.~\ref{fig:lf_market_designs} (Step 3), LFM clearing structures can be classified into two approaches according to how DN constraints are represented. The first embeds the DN model in the clearing problem: the MO receives flexibility offers and the network information needed to solve a network-constrained optimization, typically minimizing procurement cost \eqref{eq:integrated_obj} subject to voltage, thermal, and power-flow limits, as in \eqref{eq:integrated_clearing} \cite{smartnet_project}. This yields a centralized solution in which flexibility dispatch and network operation are determined simultaneously, with the DSO effectively acting as a price-taker until secure operation is restored. Constraint \eqref{eq:integrated_pf} represents the network power-flow equations, \eqref{eq:integrated_limits} enforces the operational limits of the DN, and \eqref{eq:integrated_offer} imposes the technical and operational limits of the submitted flexibility offers.
\begin{subequations}
\label{eq:integrated_clearing}
\begin{align}
\min_{\Delta P,\boldsymbol{x}^{\mathrm{DS}}}\quad
& \sum_{t\in\mathcal{T}}\sum_{a\in\mathcal{A}}
\lambda^{\mathrm{bid}}_{a,t}
\left|\Delta P_{a,t}\right|
\label{eq:integrated_obj}
\\
\text{s.t.}\qquad
& \boldsymbol{h}^{\mathrm{DS}}
\left(\boldsymbol{x}^{\mathrm{DS}}_{t},
\Delta\boldsymbol{P}_{t}\right)=\boldsymbol{0},
\qquad && \forall t\in\mathcal{T},
\label{eq:integrated_pf}
\\
& \boldsymbol{g}^{\mathrm{DS}}
\left(\boldsymbol{x}^{\mathrm{DS}}_{t},
\Delta\boldsymbol{P}_{t}\right)\leq\boldsymbol{0},
&& \forall t\in\mathcal{T},
\label{eq:integrated_limits}
\\
& \underline{\Delta P}_{a,t}
\leq \Delta P_{a,t}
\leq \overline{\Delta P}_{a,t},
 && \forall a\in\mathcal{A},\;t\in\mathcal{T}
\label{eq:integrated_offer}
\end{align}
\end{subequations}
Here, $a\in\mathcal{A}$ and $t\in\mathcal{T}$ denote the flexibility-provider and time-step indices, respectively, while $\lambda^{\mathrm{bid}}_{a,t}>0$ is the submitted offer price. $\Delta\mathbf{P}$ represents the vector of accepted flexibility quantities. \revision{The absolute-value procurement cost in \eqref{eq:integrated_obj} is appropriate when flexibility bounds are continuous and symmetric and a single offer price applies to both directions.}\footnote{\label{fn:asymm_flex}\revision{Direction-asymmetric FlexOffers, with distinct upward/downward bounds and prices, can be represented by nonnegative variables $\Delta P^{+}_{a,t}$ and $\Delta P^{-}_{a,t}$ with prices $\lambda^{+}_{a,t},\lambda^{-}_{a,t}>0$. The procurement cost then becomes $\lambda^{+}_{a,t}\Delta P^{+}_{a,t}+\lambda^{-}_{a,t}\Delta P^{-}_{a,t}$, while the net injection $\Delta P^{+}_{a,t}-\Delta P^{-}_{a,t}$ enters the network constraints.}} The vector $\mathbf{x}^{\mathrm{DS}}$ contains the DN state variables, including bus voltages, branch power flows, and other electrical quantities required by the adopted power-flow 
formulation, which may rely on a nonlinear model or suitable approximations and relaxations, e.g., the LinDistFlow. 

The second approach separates network assessment from market clearing. The DSO \revision{retains the detailed DN model,} evaluates expected operating conditions, and submits \revision{only reduced inputs}, \textit{Flexibility Requests} (volume, direction, location, and maximum price) while FSPs submit \textit{Flexibility Offers}. The MO then solves a deterministic clearing problem \eqref{eq:maximize_welfare_market} \revision{that does not embed the full DN model or a classical network-constrained OPF} \cite{prat-network-aware}. Under this structure, \eqref{eq:network_constrained_obj} maximizes social welfare and accepts offers up to the DSO willingness to pay. The approach \revision{protects privacy network data,} reduces computational burden, and yields a transparent clearing model, but may miss the optimal welfare outcome if requests are insufficient or overly conservative. Its effectiveness, therefore, depends on how the DSO quantifies the \textit{Flexibility Requests} submitted.
\begin{subequations}
\label{eq:maximize_welfare_market}
\begin{align}
\max_{\Delta\mathbf{P}^{+},\Delta\mathbf{P}^{-}}
\quad & \sum_{t\in\mathcal{T}}\sum_{z\in\mathcal{Z}}
\Big(
\lambda^{\mathrm{DSO}+}_{z,t}\Delta P^{+}_{z,t}
+
\lambda^{\mathrm{DSO}-}_{z,t}\Delta P^{-}_{z,t}
\Big)
\nonumber \\
& -
\sum_{t\in\mathcal{T}}\sum_{a\in\mathcal{A}}
\Big(
\lambda^{\mathrm{FSP}+}_{a,t}\Delta P^{+}_{a,t}
+
\lambda^{\mathrm{FSP}-}_{a,t}\Delta P^{-}_{a,t}
\Big)
\label{eq:network_constrained_obj}
\end{align}
\begin{align}
\mathrm{s.t.} & \Delta P^{+}_{z,t} = \sum_{a\in\mathcal{A}_{z}}\Delta P^{+}_{a,t},
&& \forall z\in\mathcal{Z},\;t\in\mathcal{T},
\label{eq:network_constrained_balance_up}
\\
& \Delta P^{-}_{z,t} = \sum_{a\in\mathcal{A}_{z}}\Delta P^{-}_{a,t},
&& \forall z\in\mathcal{Z},\;t\in\mathcal{T},
\label{eq:network_constrained_balance_down}
\\
& \underline{\Delta P}^{\mathrm{DSO}+}_{z,t} \leq \Delta P^{+}_{z,t} \leq
\overline{\Delta P}^{\mathrm{DSO}+}_{z,t}, && \forall z\in\mathcal{Z},\;t\in\mathcal{T},
\label{eq:network_constrained_request_up}
\\
& \underline{\Delta P}^{\mathrm{DSO}-}_{z,t} \leq \Delta P^{-}_{z,t} \leq
\overline{\Delta P}^{\mathrm{DSO}-}_{z,t}, && \forall z\in\mathcal{Z},\;t\in\mathcal{T},
\label{eq:network_constrained_request_down}
\\
& \underline{\Delta P}^{\mathrm{FSP}+}_{a,t} \leq \Delta P^{+}_{a,t} \leq
\overline{\Delta P}^{\mathrm{FSP}+}_{a,t}, && \forall a\in\mathcal{A},\;t\in\mathcal{T},
\label{eq:network_constrained_offer_up}
\\
& \underline{\Delta P}^{\mathrm{FSP}-}_{a,t} \leq \Delta P^{-}_{a,t} \leq
\overline{\Delta P}^{\mathrm{FSP}-}_{a,t}, && \forall a\in\mathcal{A},\;t\in\mathcal{T}.
\label{eq:network_constrained_offer_down}
\end{align}
\end{subequations}
Here, $\lambda^{\mathrm{DSO}\pm}_{z,t}$ and $\lambda^{\mathrm{FSP}\pm}_{a,t}$ are the nonnegative DSO and FSP prices for upward and downward flexibility, respectively, and all decision variables are nonnegative. Constraints \eqref{eq:network_constrained_balance_up}--\eqref{eq:network_constrained_balance_down} match accepted upward and downward volumes separately within each zone, while \eqref{eq:network_constrained_request_up}--\eqref{eq:network_constrained_offer_down} enforce the corresponding request and offer bounds.\footnote{\label{fn:dir_products}\revision{Upward and downward FlexRequests/FlexOffers are treated as distinct products, analogous to upward/downward frequency-regulation products. In practice, a given zone and time step typically requires only one direction, so the corresponding upward or downward subproblem can equivalently be solved on its own.}}

\section{Transformer and Cable Aging Approximation Formulation}
\label{sec:market_transformer_formulation}

This section develops a convex reformulation of the transformer and cable aging models for use in the social-welfare market-clearing framework. Building on \cite{9035437} and \cite{andrianesis2021impact}, all nonlinearities in the thermal-aging models are approximated using a piecewise-linear (PWL) formulation, which provides accurate yet computationally tractable representation.

\subsection{Transformer Degradation Formulation}
\label{subsec:aging_model}

The transformer insulation aging model follows the Arrhenius formulation in IEEE Std.~C57.91-2011 \cite{6166928}. The aging acceleration factor (FAA) is determined from the winding hottest-spot temperature (HST, $\theta^H_{\mathrm{tra},t}$) according to \eqref{eq:faa_exact}, with $\mathrm{FAA}=1$ corresponding to the nominal aging rate.

\begin{equation}
\mathrm{FAA}_{\mathrm{tra},t}
=
\exp\left(
\frac{15000}{383}
-
\frac{15000}{\theta^H_{\mathrm{tra},t}+273}
\right)
\label{eq:faa_exact}
\end{equation}

The transformer HST is determined according to \eqref{eq:theta_H}, where $\theta^{A}_{t}$ denotes the ambient temperature at the transformer location, $\Delta\theta^{TO}_{\mathrm{tra},t}$ is the top-oil temperature rise above ambient, and $\Delta\theta^{H}_{\mathrm{tra},t}$ is the winding hottest-spot temperature rise above the top-oil temperature. The top-oil temperature rise is calculated according to \eqref{eq:TO_dynamic}--\eqref{eq:TO_ultimate}, whereas the winding hottest-spot temperature rise is obtained from \eqref{eq:H_dynamic}--\eqref{eq:H_ultimate}. Both $\Delta\theta^{TO}_{\mathrm{tra},t}$ and $\Delta\theta^{H}_{\mathrm{tra},t}$ depend on the loading of the transformer.  
\begin{align}
\theta^{H}_{\mathrm{tra},t}
&= \theta^{A}_{t} +
\Delta\theta^{TO}_{\mathrm{tra},t} +
\Delta\theta^{H}_{\mathrm{tra},t},
\label{eq:theta_H}\\
\Delta\theta^{TO}_{\mathrm{tra},t}
&= \Delta\theta^{TO}_{\mathrm{tra},t-1} +
\left(\Delta\theta^{TO,U}_{\mathrm{tra},t} -
\Delta\theta^{TO}_{\mathrm{tra},t-1}
\right) \alpha^{TO}_{\mathrm{tra}},
\label{eq:TO_dynamic}\\
\Delta\theta^{TO,U}_{\mathrm{tra},t}
&=
\Delta\theta^{TO}_{R,\mathrm{tra}}
\underbrace{
\left(
\frac{K_{\mathrm{tra},t}^{2}R_{\mathrm{tra}}+1}
{R_{\mathrm{tra}}+1}
\right)^{n_{\mathrm{tra}}}
}_{f_1(K_{\mathrm{tra},t})},
\label{eq:TO_ultimate}\\
\Delta\theta^{H}_{\mathrm{tra},t}
&= \Delta\theta^{H}_{\mathrm{tra},t-1} +
\left(\Delta\theta^{H,U}_{\mathrm{tra},t}
- \Delta\theta^{H}_{\mathrm{tra},t-1}
\right)
\alpha^{H}_{\mathrm{tra}},
\label{eq:H_dynamic}\\
\Delta\theta^{H,U}_{\mathrm{tra},t} & =
\Delta\theta^{H}_{R,\mathrm{tra}}
\underbrace{K_{\mathrm{tra},t}^{2m_{\mathrm{tra}}}
}_{f_2(K_{\mathrm{tra},t})},
\label{eq:H_ultimate}\\
\alpha^{TO}_{\mathrm{tra}}
&=1-e^{-\Delta t/\tau^{TO}_{\mathrm{tra}}}, \alpha^{H}_{\mathrm{tra}}
= 1-e^{-\Delta t/\tau^{H}_{\mathrm{tra}}}.
\label{eq:H_alpha}
\end{align}
Here, the parameters $\Delta\theta^{TO}_{R,\mathrm{tra}}$,  $\Delta\theta^{H}_{R,\mathrm{tra}}$, $R_{\mathrm{tra}}$, $n_{\mathrm{tra}}$, $m_{\mathrm{tra}}$, $\tau^{TO}_{\mathrm{tra}}$, $\tau^{H}_{\mathrm{tra}}$, $\alpha^{TO}_{\mathrm{tra}}$, and $\alpha^{H}_{\mathrm{tra}}$ are transformer-specific constant parameters determined prior to the optimization. Specifically, $\Delta\theta^{TO}_{R,\mathrm{tra}}$ and $\Delta\theta^{H}_{R,\mathrm{tra}}$ denote the rated top-oil and winding hottest-spot temperature rises, $R_{\mathrm{tra}}$ is the ratio of load losses to no-load losses, $n_{\mathrm{tra}}$ and $m_{\mathrm{tra}}$ are empirically derived thermal exponents (typically 0.8), $\tau^{TO}_{\mathrm{tra}}$ and $\tau^{H}_{\mathrm{tra}}$ are the thermal time constants, while $\alpha^{TO}_{\mathrm{tra}}$ and $\alpha^{H}_{\mathrm{tra}}$ are the corresponding discrete-time thermal response coefficients \eqref{eq:H_alpha}. Finally, $K_{\mathrm{tra},t}=S_{\mathrm{tra},t}/S_{\mathrm{tra},rated}$ denotes the transformer per-unit load factor and constitutes the only time-varying quantity in \eqref{eq:TO_ultimate} and \eqref{eq:H_ultimate}.

The exponential FAA constitutes the model's first nonlinearity. The transformer thermal model also contains two further nonlinear components. The first arises from the ultimate top-oil temperature rise in \eqref{eq:TO_ultimate}, where the loading factor is raised to the empirical exponent $n_{\mathrm{tra}}$. The second originates from the ultimate winding hottest-spot temperature rise in \eqref{eq:H_ultimate}, which follows a power-law relationship with exponent $2m_{\mathrm{tra}}$. These nonlinear thermal relationships propagate to the hottest-spot temperature in \eqref{eq:theta_H} and, consequently, to the FAA in \eqref{eq:faa_exact}.

\subsection{Cable Degradation Formulation}

The cable degradation model follows the same Arrhenius-based thermal aging principle adopted for the transformer and presented in \cite{andrianesis2021impact}. Instead of the transformer winding hottest-spot temperature, the cable aging acceleration factor \eqref{eq:faa_cable_exact} is determined from the conductor core temperature, $\theta^{C}_{l,t}$,
\begin{equation}
\mathrm{FAA}_{l,t} = \exp\left(\frac{B}{363} - \frac{B}{\theta^{C}_{l,t}+273}
\right)
\label{eq:faa_cable_exact}
\end{equation}
where $B$ is a cable-specific thermal aging constant (typically around 15000). Similar to the transformer model, $\mathrm{FAA}_{l,t}=1$ corresponds to the nominal insulation aging rate at the reference conductor temperature of $90^\circ$C.

The conductor core temperature is obtained from the sum of the soil (or ambient) temperature ($\theta_t^{S/A}$), the cable surface temperature rise ($\Delta\theta_{l,t}^{CS}$), and the conductor core temperature rise ($\Delta\theta_{l,t}^{C}$), as expressed in \eqref{eq:theta_C}--\eqref{eq:C_alpha}. The corresponding thermal dynamics are identical to those presented for the transformer, where the conductor temperature evolves according to the cable loading factor, $K_{l,t}=S_{l,t}/S_{l,rated}$.
\begin{align}
\theta^{C}_{l,t} & = \theta^{S/A}_{t} +
\Delta\theta^{CS}_{l,t} + \Delta\theta^{C}_{l,t} 
\label{eq:theta_C}\\
\Delta\theta^{CS}_{l,t} & = \Delta\theta^{CS}_{l,t-1} + \left(\Delta\theta^{CS,U}_{l,t} -
\Delta\theta^{CS}_{l,t-1}
\right) \alpha^{CS}_{l},
\label{eq:CS_dynamic}\\
\Delta\theta^{CS,U}_{l,t}
&=\Delta\theta^{CS}_{R,l}
\underbrace{
K_{l,t}^{2}}_{f_3(K_{l,t})},
\label{eq:CS_ultimate}\\
\Delta\theta^{C}_{l,t} &=
\Delta\theta^{C}_{l,t-1} +
\left( \Delta\theta^{C,U}_{l,t} -
\Delta\theta^{C}_{l,t-1} \right)
\alpha^{C}_{l},
\label{eq:C_dynamic}\\
\Delta\theta^{C,U}_{l,t}
&=
\Delta\theta^{C}_{R,l}
\underbrace{
K_{l,t}^{2}
}_{f_3(K_{l,t})},
\label{eq:C_ultimate}\\
\alpha^{CS}_{l} &= 1 - e^{-\Delta t/\tau^{CS}_{l}}, \alpha^{C}_{l} = 1-e^{-\Delta t/\tau^{C}_{l}}.
\label{eq:C_alpha}
\end{align}
Here, $\Delta\theta^{CS}_{R,l}$, $\Delta\theta^{C}_{R,l}$, $\tau^{CS}_{l}$, $\tau^{C}_{l}$, $\alpha^{CS}_{l}$, and $\alpha^{C}_{l}$ denote cable-specific thermal parameters determined prior to the optimization, while $K_{l,t}$ represents the cable per-unit loading. 

Similar to the transformer degradation model, the cable formulation contains three nonlinear components. The first is the exponential Arrhenius aging function in \eqref{eq:faa_cable_exact}. The remaining two nonlinearities arise from the quadratic dependence of the ultimate cable surface temperature rise \eqref{eq:CS_ultimate} and the ultimate conductor temperature rise \eqref{eq:C_ultimate} on the cable loading factor. These nonlinear thermal relationships propagate to the conductor core temperature in \eqref{eq:theta_C} and consequently to the aging acceleration factor in \eqref{eq:faa_cable_exact}. 

\subsection{Piecewise Linearization and Epigraph Reformulation}

\subsubsection{Aging Factor and Thermal-Rise Piecewise Linearization}

The FAA in~\eqref{eq:faa_exact} and \eqref{eq:faa_cable_exact} is approximated using a PWL representation over the operating temperature range
$[\theta_{\min},\theta_{\max}]$. The approximation \eqref{eq:pwl_faa} is constructed from $M$ linear chord segments connecting consecutive temperature breakpoints.
\begin{equation}
\tilde{f}^{FAA}
= a_k^{\mathrm{FAA}}\theta+b_k^{\mathrm{FAA}},
\theta_{k-1}\le\theta<\theta_k,
k=1,\ldots,M
\label{eq:pwl_faa}
\end{equation}
The coefficients of each segment are computed from two consecutive FAA values, $a_k^{\mathrm{FAA}}=\frac{F_{k}^{\mathrm{FAA}}-F_{k-1}^{\mathrm{FAA}}}{\theta_k-\theta_{k-1}}$, $b_k^{\mathrm{FAA}}=F_{k-1}^{\mathrm{FAA}} - a_k^{\mathrm{FAA}}\theta_{k-1}$.

For the thermal rise functions, in contrast to \cite{9035437} and \cite{andrianesis2021impact}, which used first-order Taylor approximations, the proposed method employs PWL approximations, providing higher accuracy, as validated in Section~\ref{sec:tra_cable_reform_accur}. Let
$f_r(K)\in\{f_1(K_{\mathrm{tra}}),f_2(K_{\mathrm{tra}}),f_3(K_{\mathrm{c}})\}$ denote any of the nonlinear thermal rise functions.
Each function is approximated over $M$ loading segments \eqref{eq:pwl_general}.
\begin{equation}
\tilde f_r(K)
= a^{r}_{k}K+b^{r}_{k},
K_{k-1}\le K<K_k,
k=1,\ldots,M
\label{eq:pwl_general}
\end{equation}
Here, $a^{r}_{k}$ and $b^{r}_{k}$ are the slope and intercept of the $k$-th segment for function $f_r(\cdot)$. 

\subsubsection{Epigraph Reformulation}
\label{sec:epigraph_reformulation}
The nonlinear functions~\eqref{eq:faa_exact}, \eqref{eq:TO_ultimate}, \eqref{eq:H_ultimate}, \eqref{eq:faa_cable_exact}, \eqref{eq:CS_ultimate}, and \eqref{eq:C_ultimate}, can be embedded in the optimization through \emph{epigraph} constraints rather than nonlinear equalities. For a scalar
function $g(x)$, the epigraph is the set
$\{(x,y):y\ge g(x)\}$. When $g(\cdot)$ is convex and the objective minimizes
$y$, the optimal solution satisfies $y^\star=g(x^\star)$ \cite{boyd2004convex}, i.e., the auxiliary variable is pushed to its minimum feasible value. Consequently, each piecewise segment in \eqref{eq:pwl_faa} and \eqref{eq:pwl_general} is represented by one linear inequality ($y \geq a_k x+b_k$).

The function $\tilde{f}^{\mathrm{FAA}}$ is strictly convex for all physically relevant temperatures. Moreover, the thermal rise functions $f_1(K)$ and $f_2(K)$ are strictly convex for typical transformer parameters $n=m=0.8$~\cite{6166928} and $R_{\mathrm{tra}}>0$, while $f_3(K)$ is convex as it is quadratic in $K$. Since the FAA is monotonically increasing with these thermal rise functions, minimizing $\tilde{f}^{\mathrm{FAA}}_t$ drives their epigraph variables to their minimum feasible values. Consequently, all corresponding epigraph constraints are binding at optimality, recovering the original nonlinear equalities. A formal proof for the transformer is provided in Appendix~\ref{app:conv_monot_proof}. The cable proof is omitted for brevity, as it follows analogously from the transformer case.

\section{Network Benefit Monetization and Local Market Clearing}
\label{sec:DSO_Network_benefits}

The valuation framework is defined relative to a baseline operating point with no flexibility procurement under the anticipated loading profile. From this reference, the DSO quantifies avoided impacts of flexibility activation, transformer and cable aging, MV/LV losses, and voltage violations, which determine the corresponding willingness to pay.
\revision{Additional value streams could also be incorporated into the DSO willingness-to-pay, for example CAPEX deferral, reliability improvements, energy not served, or additional contractual activation fees beyond the bid; extending the valuation in this direction is left for future work.}

\subsection{Transformer and Cable Aging Cost}
The baseline transformer and cable aging costs are first established. Based on the initial operating point, the corresponding baseline aging acceleration factors, $\mathrm{FAA}^{\mathrm{base}}_{\mathrm{tra},t}$ and $\mathrm{FAA}^{\mathrm{base}}_{l,t}$, are calculated for all transformers and cables within the feeder considered in the market-clearing algorithm. Then, the baseline FAA can be converted into a monetary aging cost by using the normal insulation lifetime. Let $L$ denote the normal transformer and cable lifetime in hours and $c^{\mathrm{rep}}$ the corresponding replacement cost. For each time interval $t$, the baseline aging cost is given in \eqref{eq:baseline_aging_cost} with respect to the loss of life (LoL). The corresponding aging benefit is then defined in \eqref{eq:aging_benefit}, which represents the monetary reduction in aging achieved through flexibility activation. Here, $C^{\mathrm{opt}}_{\{\mathrm{tra},l\},t}$ denotes the optimized aging cost for the transformer or cable after the procurement of flexibility.
\begin{align}
\label{eq:baseline_aging_cost}
C^{\mathrm{base}}_{\{\mathrm{tra},l\},t}
& = c^{\mathrm{rep}}_{\{\mathrm{tra},l\}}
\underbrace{\frac{FAA^{\mathrm{base}}_{\{\mathrm{tra},l\},t}} {L_{\{\mathrm{tra},l\}}}\Delta t
}_{\mathrm{LoL}^{\mathrm{base}}_{\{\mathrm{tra},l\},t}}, && \forall t\in\mathcal{T} \\
B^{\mathrm{age}}_{\{\mathrm{tra},l\},t}
&= C^{\mathrm{base}}_{\{\mathrm{tra},l\},t}
- C^{\mathrm{opt}}_{\{\mathrm{tra},l\},t}, && \forall t\in\mathcal{T}
\label{eq:aging_benefit}    
\end{align}

The loading factor of each transformer or cable is expressed in per-unit form using \eqref{eq:Ku_exact} based on the nominal apparent power. Although \eqref{eq:Ku_exact} provides the exact loading level, it is nonlinear with respect to the active-power variation introduced by flexibility and is therefore not directly suitable for the market-clearing formulation. To obtain a tractable representation, the loading factor is linearized around the baseline operating point $(P_t^{\mathrm{base}},Q_t^{\mathrm{base}})$ by means of a first-order Taylor approximation. Assuming that the reactive-power variation remains unaffected over the market-clearing horizon, the resulting approximation is written in \eqref{eq:Ku_taylor}, where $K_{\{tra,l\},t}^{\mathrm{base}}$ is the baseline loading factor and $\Delta P_t$ is the flexibility-induced active-power change. Therefore, flexibility modifies the loading factor via a linear sensitivity term, enabling the updated thermal state, aging acceleration factor, and aging cost to be efficiently embedded in the market-clearing optimization.
\begin{align}
\label{eq:Ku_exact}
& K_{\{tra,l\},t}=\frac{\sqrt{P_t^2+Q_t^2}}{S_n}, \\
& K_{\{tra,l\},t} \approx K_{\{tra,l\},t}^{\mathrm{base}} + \frac{\partial K_{\{tra,l\},t}}{\partial P_t}\Big|_{\mathrm{base}} \Delta P_{\{tra,l\},t},
\label{eq:Ku_taylor} \\
&\frac{\partial K_{\{tra,l\},t}}{\partial P_t}\Big|_{\mathrm{base}}
= \frac{P_t^{\mathrm{base}}}{S_n\sqrt{(P_t^{\mathrm{base}})^2+(Q_t^{\mathrm{base}})^2}}.
\label{eq:Ku_sensitivity}
\end{align}

\subsection{Loss Cost Benefit Estimation}
\label{sec:losses_cost}
Beyond aging, flexibility also modifies power flows and thus network losses, which may raise or lower DSO operating costs relative to the baseline. The corresponding economic impact is quantified in \eqref{eq:loss_impact_cost} using the wholesale energy price $\lambda^W_t$.
\begin{equation}
B_{loss,t} = C_{loss,t}^{base} - C_{loss,t}^{opt} = \lambda_t^{W} \left( P_{loss,t}^{base} - P_{loss,t}^{opt} \right) \Delta t
\label{eq:loss_impact_cost}
\end{equation}

\subsubsection{MV Losses, Sensitivity-Based Coefficients}
\label{sec:mv_loss_sensitivity}

Since no network model is incorporated into the optimization-clearing algorithm, calculating the reduction in network losses directly is not feasible. For this purpose, sensitivity calculations from baseline operation are used. More specifically, during the DSO internal process, the total losses can be computed from the complex bus-voltage state, since the latter fully determines the branch currents and complex power flows in the network. Let $\mathbf{v}_t$ denote the complex bus-voltage vector at time step $t$. Based on the network branch admittance, the branch currents ($\mathbf{i}_t$) at the sending and receiving ends are given by \eqref{eq:branch_currents}. The corresponding complex power flows ($\textbf{s}_t$) are then computed in \eqref{eq:complex_power_flows}. Accordingly, the total active-power losses are obtained in \eqref{eq:losses_calculation}. Here, $\Re(\cdot)$ represents the real part of a complex quantity, $\odot$ the element-wise product, and $\mathcal{E}$ the branch set. \revision{The subscripts $f$ and $t$ on the branch quantities denote the from (sending) and to (receiving) ends of each network branch, respectively.}
\begin{align}
\label{eq:branch_currents}
\mathbf{i}_{f,t} &= \mathbf{Y}_{f}\mathbf{v}_t,
&&\mathbf{i}_{t,t} = \mathbf{Y}_{t}\mathbf{v}_t, \\
\label{eq:complex_power_flows}
\mathbf{s}_{f,t} &= \mathbf{v}_{f,t} \odot \mathbf{i}_{f,t}^{*}, &&\mathbf{s}_{t,t} = \mathbf{v}_{t,t} \odot \mathbf{i}_{t,t}^{*},
\end{align}
\begin{equation}
\label{eq:losses_calculation}
L_t = \sum_{\ell \in \mathcal{E}} \Re \left( s_{f,\ell,t} + s_{t,\ell,t} \right),
\end{equation}

Although the quantity of interest is the sensitivity of losses with respect to an active-power variation at bus $j$, the losses are evaluated through the intermediate voltage state. Thus, total losses can be expressed as $L_t = L(\boldsymbol{\phi}_t, \mathbf{V}_t)$, and their sensitivity to a nodal active-power injection is derived by applying the chain rule, as shown in \eqref{eq:losses_chain_rule},
\begin{align}
\label{eq:losses_chain_rule}
&S^{\mathrm{loss}}_{t,j}
\equiv  \frac{\partial L_t}{\partial P_{j,t}}
\!= \!\!\!\!\!
\sum_{i \in \mathcal{PV}\cup\mathcal{PQ}} \!\!\!
\frac{\partial L_t}{\partial \phi_{i,t}}
\frac{\partial \phi_{i,t}}{\partial P_{j,t}}
+ \!\!\!\sum_{i \in \mathcal{PQ}} \!\!\frac{\partial L_t}{\partial V_{i,t}} \frac{\partial V_{i,t}}{\partial P_{j,t}}
\end{align}
where $\mathcal{PV},\mathcal{PQ}$ are the sets of PV and PQ buses.

The first set of derivatives, i.e., \(\partial L_t/\partial \phi_{i,t}\) and \(\partial L_t/\partial V_{i,t}\), is obtained numerically by perturbing the voltage angles ($\phi_i$) and magnitudes ($V_i$) around the baseline operating point and re-evaluating the total losses through the branch-current and branch-power expressions above. The second set of derivatives, i.e., \(\partial \phi_{i,t}/\partial P_{j,t}\) and \(\partial V_{i,t}/\partial P_{j,t}\), is obtained from the AC power-flow Jacobian ($\mathbf{J_t}$) at the baseline operating point.  Solving this linear system yields the sensitivity of the total losses to a unit active-power perturbation at bus $j$. 

Once these coefficients have been obtained for all buses and time steps, the variation of total active-power losses due to flexibility activation can be approximated through a linear expression. Hence, by multiplying the time-varying total-loss sensitivities by the corresponding active-power adjustments, the DSO can estimate whether flexibility increases or reduces total losses, which can subsequently be monetized in \eqref{eq:mv_approx_losses_cost_impact}.
\begin{align}
\label{eq:mv_approx_losses_cost_impact}
B^{MV}_{loss,t} &\approx \lambda_t^W \sum_{j} \left(S^{\mathrm{loss}}_{t,j}\,\Delta P_{j,t} \right) \Delta_t
\end{align}

\subsubsection{LV Feeders, Equivalent Loss-Intensity Factor}
\label{sec:lv_loss_factor}
For LV feeders, a detailed analytical network model is generally unavailable to the DSO, as LV circuits are frequently modified after commissioning, while topology and conductor data are often incomplete in operational databases. In this case, the loss sensitivity at the secondary transformer terminal is replaced by a simplified feeder-level approximation \eqref{eq:lv_losses}. A first-order Taylor expansion around the baseline operating point yields a linear approximation, which is used as the time-varying LV loss sensitivity in the market-clearing model \eqref{eq:lv_sensitivity_losses}. The coefficient $k_s$ depends on the electrical characteristics and geographical extent of the LV feeder, and therefore varies. In practice, for typical LV feeders, peak-loss ratios generally range from \revision{4\% to 9\% \cite{strbac2014management}}, with compact urban feeders at the lower end and long rural feeders at the upper end.
\begin{align}
P^{\mathrm{LV,loss}}_{s,t} & \approx k_s P_{s,t}^2,
\label{eq:lv_losses} \\
\label{eq:lv_sensitivity_losses}
\Delta P^{\mathrm{LV,loss}}_{s,t} & \approx 2k_s P^{\mathrm{base}}_{s,t}\,\Delta P_{s,t},
\end{align}
Here, $P_{s,t}$ is the total active power supplied through the corresponding secondary transformer ($s$) and $k_s$ is a feeder-dependent loss coefficient. Therefore, the cost impact of LV losses can be calculated for the new operating point using \eqref{eq:lv_loss_impact_cost}.
\begin{equation}
B^{LV}_{loss,s,t} = \lambda_t^W \Delta P^{\mathrm{LV,loss}}_{s,t} \Delta t.
\label{eq:lv_loss_impact_cost}
\end{equation}

\subsection{Voltage Violation Modeling}

The voltage-related component is introduced to quantify the economic value of flexibility for alleviating undervoltage and overvoltage violations. Using post-simulation analysis, if voltage violations are identified in the baseline condition, the proposed market-clearing model uses voltage sensitivity coefficients to estimate how the activation of flexibility offers modifies the nodal voltages. In this way, the market can select the most cost-effective combination of flexibility offers that reduces the existing voltage violations. The baseline undervoltage and overvoltage violations \eqref{eq:voltage_violations_baseline} are first calculated as fixed parameters before the market-clearing problem is solved,
\begin{equation}
\label{eq:voltage_violations_baseline}
\!\!s^{UV,0}_{i,t} = \max\left(0,\underline{V}-V^{0}_{i,t}\right),
s^{OV,0}_{i,t} = \max\left(0,V^{0}_{i,t}-\overline{V}\right) 
\end{equation}
where the superscript $(\cdot)^0$ denotes a quantity evaluated at the baseline operating point before flexibility activation. $V^0_{i,t}$ is the voltage magnitude, while $\underline{V}$ and $\overline{V}$ denote the lower and upper voltage limits, respectively. The parameters $s^{UV,0}_{i,t}$ and $s^{OV,0}_{i,t}$ therefore represent the anticipated voltage violation that the DSO aims to mitigate through flexibility procurement.

Given these baseline violations, the clearing model next approximates how flexibility changes the nodal voltages. Voltage sensitivity coefficients $S^P_{i,a,t}$ quantify the incremental change in voltage magnitude at bus $i$ due to an active-power variation from provider $a$ ($\Delta P_{a,t}$), and are obtained from the inverse of the AC power-flow Jacobian at the baseline operating point, $\mathbf{J}_t^{-1,\mathrm{base}}[\Delta\mathbf{P}_t^{\!\top},\mathbf{0}^{\!\top}]^{\!\top}$. The post-activation voltage is then written as the baseline voltage plus the linear contribution of all accepted offers \eqref{eq:voltage_sensitivity}.
Any remaining undervoltage or overvoltage after flexibility activation is captured by non-negative slack variables in \eqref{eq:undervoltage_overvoltage_slack}--\eqref{eq:nonnegative_slack}: if $s^{UV}_{i,t}$ or $s^{OV}_{i,t}$ is positive, the corresponding limit is still violated; if both are zero, the voltage at bus $i$ lies within $[\underline{V},\overline{V}]$.
\begin{align}
\label{eq:voltage_sensitivity}
V_{i,t}& = V^{0}_{i,t} + \sum_{a\in\mathcal{A}} S^P_{i,a,t}\Delta P_{a,t} \\
\label{eq:undervoltage_overvoltage_slack}
V_{i,t} &\ge \underline{V} - s^{UV}_{i,t},
V_{i,t} \le \overline{V} + s^{OV}_{i,t}, \\
\label{eq:nonnegative_slack}
s^{UV}_{i,t}&,\,s^{OV}_{i,t} \ge0.
\end{align}

Finally, the voltage benefit monetizes the \emph{reduction} of violations relative to the baseline \eqref{eq:voltage_objective}. The terms $\bigl(s^{UV,0}_{i,t}-s^{UV}_{i,t}\bigr)$ and $\bigl(s^{OV,0}_{i,t}-s^{OV}_{i,t}\bigr)$ are the mitigated undervoltage and overvoltage at bus $i$ (in~p.u.), weighted by the coefficients $C^{UV}_{i,t}$ and $C^{OV}_{i,t}$ (\euro/p.u.). If violations are fully eliminated ($s^{UV}_{i,t}=s^{OV}_{i,t}=0$), no further voltage-driven flexibility is valued at that bus. \revision{The $C^{UV}_{i,t}$ and $C^{OV}_{i,t}$ coefficients admit two interpretations. They may be set large so that $B^{\mathrm{V}}$ acts as an operational soft penalty and congestion is relieved whenever feasible offers exist. Alternatively, they may encode a genuine DSO willingness-to-pay obtained from longer-term studies that monetize, for example, reinforcement deferral or the value of lost load associated with sustained violations. In the present work, $C^{UV}$ and $C^{OV}$ are used in the soft-penalty sense, since no network reinforcement is assumed, residual violations should be mitigated, so a large $C$ is adopted.}
\begin{equation}
\label{eq:voltage_objective}
B^{\mathrm{V}}_t \! = \! \sum_{i\in\mathcal{N}} C^{UV}_{i,t}
\left(s^{UV,0}_{i,t} - s^{UV}_{i,t} \right)
+ C^{OV}_{i,t} \left(s^{OV,0}_{i,t} - s^{OV}_{i,t}
\right)
\end{equation}

\subsection{Complete Local Market-Clearing Algorithm}

The components above are now assembled into a complete clearing algorithm. In contrast to fixed flexibility requests, e.g., \cite{prat-network-aware}, the proposed \textit{dynamic flexibility request} does not prescribe a predefined volume: the DSO's willingness to procure is endogenous, obtained by embedding the aggregate economic impact of each activation directly in the clearing objective. Analogous to economic dispatch, where generators submit cost coefficients $(a_g,b_g,c_g)$ and the operator clears subject to network constraints, the DSO publishes transformer and cable characteristics together with sensitivity matrices at the anticipated operating point, while an independent market operator matches the bids. The resulting DSO benefit aggregates the aging terms of the primary and secondary transformers and MV cables, the MV and LV loss benefits, and the voltage-violation component, all relative to the anticipated baseline operating point~\eqref{eq:dso_total_benefit}.
\begin{equation}
B_t^{DSO} = B_{tra,t}^{age} + B_{l,t}^{age} + B^{MV}_{loss,t} + B^{LV}_{loss,t} + B^V_t
\label{eq:dso_total_benefit}
\end{equation}
The DSO clears flexibility by maximizing net welfare, where inequality constraints are used for all piecewise-linearizations \eqref{eq:market_obj_final}-\eqref{eq:market_ineq_constraints_2}.
\begin{align}
\max \quad &
\sum_{t\in\mathcal{T}}
\left(B^{\mathrm{DSO}}_t - \sum_{a\in\mathcal{A}}
\lambda^{\mathrm{FSP}}_{a,t}\, {\Delta P}^{\mathrm{flex}}_{a,t}\,
\Delta t
\right)
\label{eq:market_obj_final}
\end{align}
\begin{align}
&\underline{\Delta P}^{\mathrm{flex}}_{a,t} \le \Delta P^{\mathrm{flex}}_{a,t} \le \overline{\Delta P}^{\mathrm{flex}}_{a,t},
&& \forall a\in\mathcal{A},\forall t\in\mathcal{T},
\label{eq:bid_bounds} \\
&\Delta P^{\mathrm{flex,sec}}_{s,t} = \sum_{\substack{a\in\mathcal{A}| \sigma(a)=s}}
\Delta P^{\mathrm{flex}}_{a,t}, &&
\forall s\in\mathcal{S}, \forall t\in\mathcal{T},
\label{eq:sec_flex_agg} \\
&\Delta P^{\mathrm{flex,pri}}_{p,t} =
\sum_{\substack{s\in\mathcal{S} |\pi(s)=p}} \Delta P^{\mathrm{flex,sec}}_{s,t}, &&
\forall p\in\mathcal{P}, \forall t\in\mathcal{T},
\label{eq:pri_flex_agg} \\
& \eqref{eq:theta_H}-\eqref{eq:H_alpha}, 
\eqref{eq:theta_C}-\eqref{eq:aging_benefit}, \eqref{eq:Ku_taylor}-\eqref{eq:Ku_sensitivity} && \forall t\in\mathcal{T} 
\label{eq:market_ineq_constraints_1}\\
& \eqref{eq:mv_approx_losses_cost_impact}, \eqref{eq:lv_sensitivity_losses}-\eqref{eq:lv_loss_impact_cost}, \eqref{eq:voltage_sensitivity}-\eqref{eq:voltage_objective}  && \forall t\in\mathcal{T}
\label{eq:market_ineq_constraints_2}
\end{align}
Here, $\sum_{a}\lambda_{a,t}^{\mathrm{FSP}}\,\Delta P_{a,t}^{\mathrm{flex}}$ is the total FSP payment at bid prices $\lambda_{a,t}^{\mathrm{FSP}}$ and dispatched flexibility $\Delta P_{a,t}^{\mathrm{flex}}$. The mapping $\sigma(a)$ associates flexibility provider $a$ with its corresponding secondary transformer $s$, while $\pi(s)$ maps each secondary transformer $s$ to its upstream primary transformer $p$. The bidding structure to determine FlexOffers is beyond the scope of this article.
\section{Case Study}
\label{sec:case_study}

The numerical experiments use the modified CIGRE MV network\revision{, adopted as a representative and widely used distribution benchmark to ensure a transparent numerical setting and to facilitate fair comparison with existing and future LFM studies under shared open inputs}, together with base- and stress-load scenarios, FSP offers, wholesale prices, and transformer/cable thermal--economic parameters provided in the publicly available Zenodo repository~\cite{panagi2026zenodo}, which enables reproducibility and reuse in future studies.

\revision{Two operating scenarios are considered. The stress case retains the base-case demand profile and superimposes uncontrolled residential EV charging corresponding to a $30\%$ EV ownership share at each secondary substation. This penetration level is selected to provide a meaningful stress condition for evaluating the proposed methodology.}
For a comprehensive understanding of the non-flexibility activation case, readers are referred to \cite{panagi2026zenodo}.
Clearing-specific coefficients not prescribed therein are set as follows: voltage limits $\underline{V}=0.90$~p.u.\ and $\overline{V}=1.10$~p.u.\ are enforced via soft penalties with $C^{UV}_{i,t}=C^{OV}_{i,t}=\SI{10000}{EUR/pu}$ \revision{(the rationale and the effect of this choice are discussed in Section~\ref{sec:network_approx})}, a uniform peak-loss ratio of $4\%$ is assumed for all LV feeders \cite{strbac2014management}, and baseline loads are modeled with a constant power factor of 0.95. Only active-power flexibility is considered in the market clearing. The baseline operating point is obtained from a full AC power-flow analysis, while reactive power injections remain fixed during optimization. The framework can be extended to reactive-power flexibility by introducing the corresponding decision variables and sensitivities; however, this is outside the scope of this work. 

\revision{For comparison, the network-aware flexibility-request method of~\cite{prat-network-aware} is implemented using the same network model, anticipated baseline operating points, FlexOffers, and voltage limits as the proposed framework, in both bus-specific and five-zone configurations. Consistent with that reference, flexibility needs are first quantified by a LinDistFlow FlexRequest OPF and then cleared by the deterministic zonal request--offer model. Although the original FlexRequest formulation can incorporate forecast uncertainty through chance constraints over a family of expected operating conditions, the present comparison uses its deterministic counterpart evaluated at the same single anticipated baseline as the proposed method. This choice enables a like-for-like comparison focused on methodological differences in flexibility-request formulation and clearing, rather than on the treatment of forecast uncertainty.} 

\revision{All reported market-clearing problems are solved using the Gurobi solver~\cite{gurobi}. Because the base and stress instances share the same decision structure and differ only in their exogenous data, they have identical problem dimensions. For the CIGRE case study, considering a $24$~h horizon, the resulting model comprises approximately $8.6{\times}10^{3}$ variables and $1.8{\times}10^{5}$ constraints. Each transformer and cable thermal model comprises approximately $194$ variables, plus $720$ variables associated with all bidders and $72$ variables per bus. The large majority of the constraints arise from the piecewise-linear epigraph approximations of the transformer and cable thermal/aging models. On a laptop equipped with an M4 Max processor ($14$ cores, $36$~GB memory), a full market-clearing instance is solved in approximately \SI{0.68}{\second}.}
\section{Results and Discussion}
\label{sec:Results_Discussions}

\begin{table}[b!]
\centering
\caption{Accuracy of the linearized cable and transformer aging models for a single day simulation.}
\label{tab:aging_accuracy}
\renewcommand{\arraystretch}{1}
\setlength{\tabcolsep}{3pt}
\begin{tabular}{lccccc}
\toprule
Model & FEQA & Cost [€]& $\|\mathrm{FAA}\|_\infty$ & $\|\theta\|_\infty$ 
& RMSE$(\theta)$ 
\\
\midrule
\multicolumn{6}{c}{\textbf{Transformer}}\\
\midrule
Exact      & 1.338 & 446.14 & -- & -- & -- \\
Model \cite{9035437}  & 1.249 & 416.47 & 0.380 & 1.736 & 0.873 \\
Proposed   & 1.340 & 446.54 & 0.004 & 0.002 & 0.001 \\
\midrule
\multicolumn{6}{c}{\textbf{Cable}}\\
\midrule
Exact     & 0.148 & 1.831 & --    & --    & -- \\
Model \cite{andrianesis2021impact} & 0.125 & 1.542 & 0.230 & 5.779 & 2.595 \\
Proposed  & 0.148 & 1.834 & 0.002 & 0.005 & 0.003 \\
\bottomrule
\end{tabular}
\end{table}

\subsection{Transformer and Cable Aging Formulation Accuracy}
\label{sec:tra_cable_reform_accur}
The proposed transformer and cable aging formulations are validated and compared against the existing approaches of \cite{andrianesis2021impact,9035437}. The comparison is carried out under the base-case operating scenario using the primary substation transformer between buses 0--1 and the cable connecting buses 1--2. The objective is to assess both the approximation accuracy and the resulting impact on the estimated aging cost.

\begin{figure}
    \centering
    \includegraphics[width=1\linewidth]{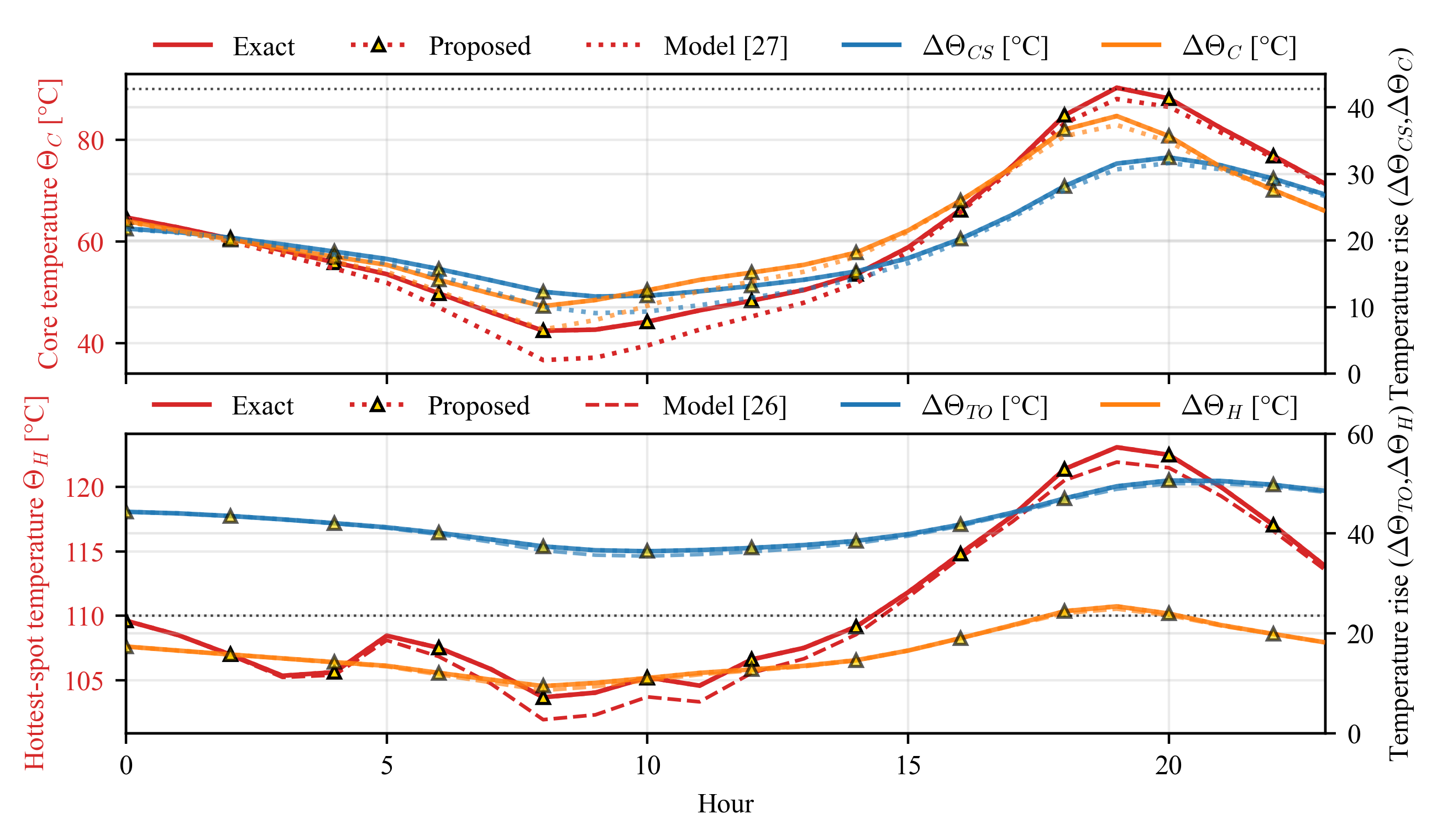}
    \caption{Comparison of the exact, existing linearized, and proposed thermal models for the cable and transformer in terms of conductor/hot-spot temperature and the corresponding temperature-rise components.}
    \label{fig:thermal_models_comparison}
\end{figure}

The thermal-model validation results are presented in Fig.~\ref{fig:thermal_models_comparison}, while the corresponding quantitative accuracy metrics are summarized in Table~\ref{tab:aging_accuracy}. It can be observed that Taylor-based approximations introduce deviations in the temperature-rise components, which propagate to the transformer hot-spot and cable-core temperatures, ultimately affecting the estimated aging cost. The equivalent aging factor (FEQA), defined as the 24~h time-weighted average of the aging acceleration factor, for the proposed model reproduces the exact value, whereas the benchmark linear model deviates more notably, indicating sub-nominal thermal aging over the day.

For the transformer, Model~\cite{9035437} exhibits a maximum temperature error of \SI{1.74}{\celsius} and underestimates the daily aging cost by approximately \EUR30 (6.7\%), whereas the proposed formulation reduces the maximum error to only \SI{0.002}{\celsius} and yields an aging cost almost identical to that of the exact model.
Similarly, for the cable, the proposed formulation achieves a near-zero maximum temperature error, while Model~\cite{andrianesis2021impact} reaches a maximum deviation of \SI{5.8}{\celsius} and an RMSE of \SI{2.6}{\celsius}. Consequently, the proposed model provides an almost identical daily aging estimate, whereas Model~\cite{andrianesis2021impact} underestimates the aging cost by approximately 16\%.

\subsection{Base Case}

\subsubsection{Distributed Locational Marginal Prices}

\begin{figure}
    \centering
    \includegraphics[width=1\linewidth]{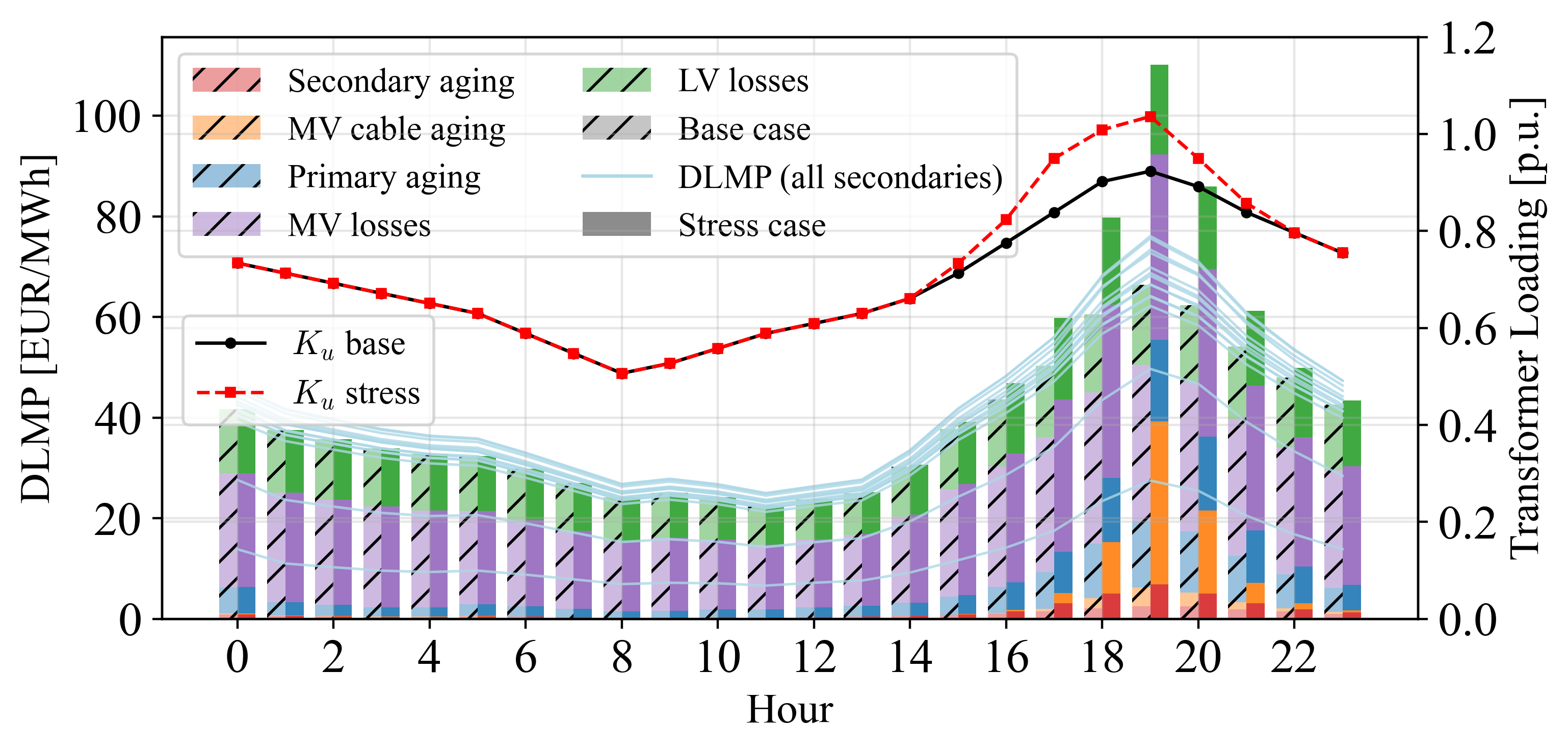}
    \caption{Hourly decomposition of the proposed DSO willingness-to-pay (DLMP) for secondary transformer~1 (between buses~3 and~15) under the base-case and stress-case operating conditions. Light-blue curves represent the base-case DLMP profiles of all secondary substations.}
    \label{fig:trafo2_split_lmp}
\end{figure}

Before the market-clearing process, an \textit{ex-ante} analysis of the distributed locational marginal prices (DLMP) is conducted. \revision{This indicative analysis is obtained from small load perturbations at each secondary substation around the anticipated baseline and is used only to illustrate the magnitude and locational variation of the DSO willingness-to-pay.} Figure~\ref{fig:trafo2_split_lmp} illustrates the decomposition of the DLMP for a representative secondary substation. It can be observed that network losses constitute the dominant component of the flexibility value throughout the day. The contributions of secondary transformer aging and MV cable aging are considerably smaller due to the relatively low loading of these assets, contributing to approximately \SI{10}{EUR/MWh} during peak loading periods. Furthermore, the contribution of primary transformer aging increases significantly during the evening peak, following the corresponding increase in transformer loading.

The light-blue curves in Fig.~\ref{fig:trafo2_split_lmp} present the resulting DLMPs for all fourteen secondary substations. A clear spatial variability can be observed. As described in Section~\ref{sec:DSO_Network_benefits}, the DSO's willingness to procure flexibility varies across the network, since all cost components, except the primary transformer aging, are location-dependent. Consequently, each secondary substation is associated with a different DLMP reflecting its local network conditions and operational constraints.

\subsubsection{Market Clearing Results} 

\begin{figure}
    \centering
    \includegraphics[width=0.97\linewidth]{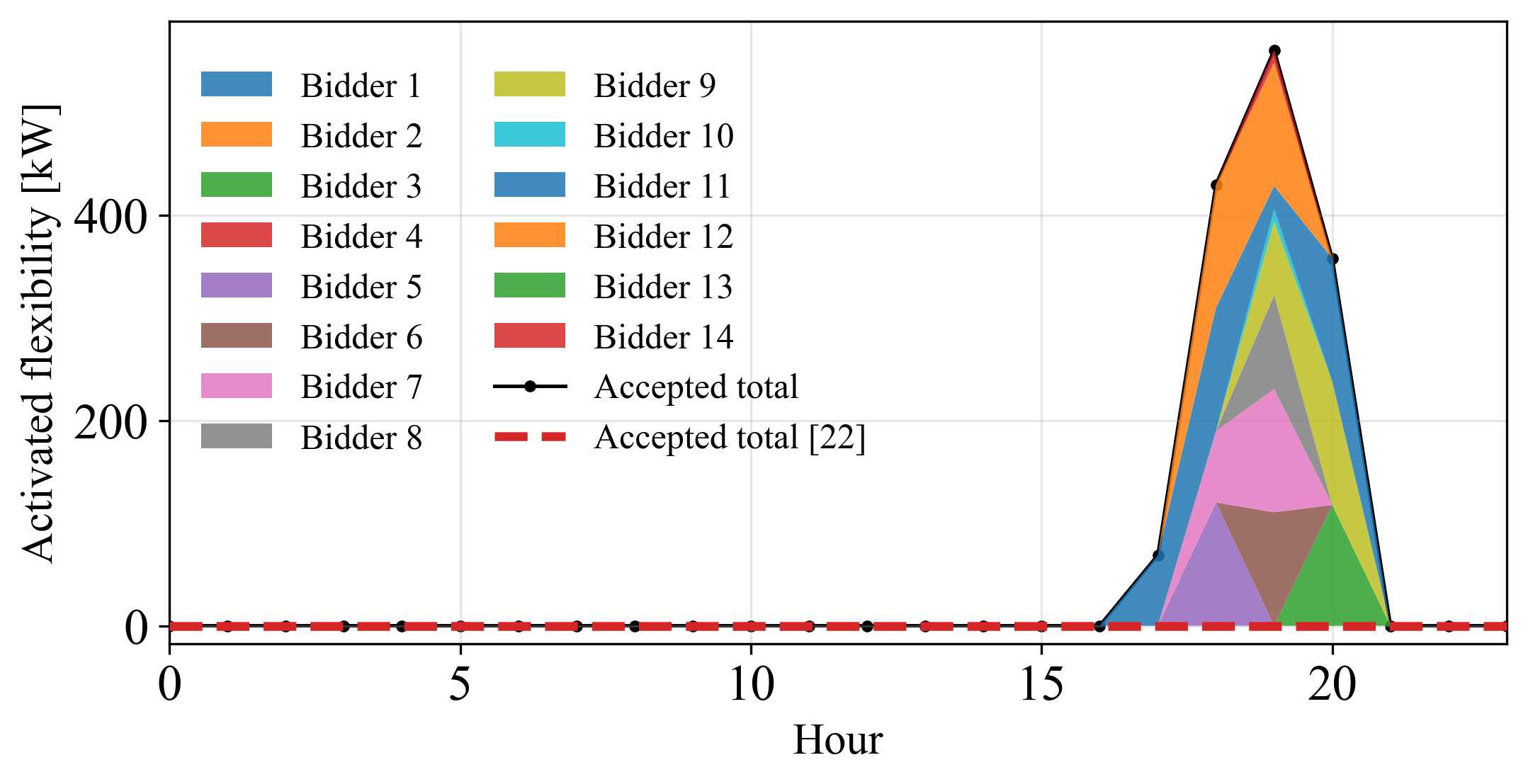}
    \caption{Hourly activated flexibility under the proposed methodology. Each bidder corresponds to a single secondary substation. The red dashed line shows the total accepted flexibility obtained using the benchmark methodology of \cite{prat-network-aware}.}
    \label{fig:market_results_base_case}
\end{figure}

Market-clearing results for the base-case scenario are presented in Fig.~\ref{fig:market_results_base_case} and Table~\ref{tab:market_results}. As expected from the DLMP analysis, flexibility activation increases significantly during the late-afternoon and evening hours, when transformer loading, voltage violations, and network losses become more critical. 

Comparing the proposed framework with the static flexibility-request methodology of \cite{prat-network-aware}, Fig.~\ref{fig:market_results_base_case} shows that the benchmark does not activate any flexibility. This is because it focuses exclusively on congestion management, while the adopted LinDistFlow approximation fails to identify the existing voltage and loading congestion. Consequently, no flexibility request is issued despite the network operating close to its limits. This also highlights the limited-liquidity issue noted in \cite{khorasany2020framework}, where the local flexibility market remains inactive for most hours, preventing flexibility from providing additional operational and economic benefits to the DSO.

\begin{figure}
    \centering
    \includegraphics[width=0.97\linewidth]{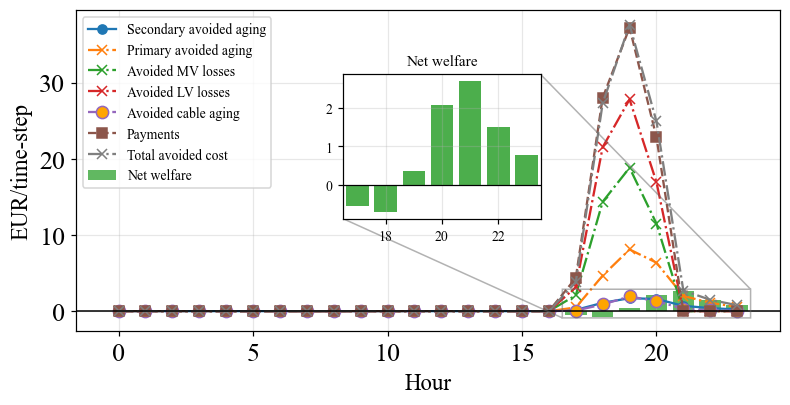}
    \caption{Temporal breakdown of the market-clearing  benefits obtained from flexibility activation in the base-case scenario.}
    \label{fig:breakdown_base_results}
\end{figure}

Fig.~\ref{fig:breakdown_base_results} further decomposes the DSO economic benefit from the clearing algorithm into its individual components. While a slightly negative net welfare is observed during the first activation intervals (17:00--18:00), this investment is compensated by the considerably higher avoided costs during the evening peak (19:00--23:00), demonstrating the value of considering inter-temporal asset degradation within the market-clearing process. The benefit breakdown also provides useful insight into the economic drivers of flexibility procurement. 

\setlength{\tabcolsep}{2pt}
\begin{table}
\caption{Summary of market-clearing results under base case and stress case operating conditions.}
\label{tab:market_results}
\centering
\begin{tabular}{lcc}
\hline
\textbf{Metric} & \textbf{Base case} & \textbf{Stress case} \\
\hline
Total energy [\si{MWh/day}]     & 808.6 & 811.3 [+0.3\%] \\
Peak network load [\si{MW}]     &  44.3 &  44.9 [+1.4\%] \\
Peak hour & {20:00--21:00}  & {20:00--21:00}  \\
Total activated flexibility [kWh] & 1417 / 0 / 0 & 2720 / 391 / 459 \\
Maximum activated flexibility [kW] & 560 / 0 / 0 & 592 / 227 / 256 \\
Avoided primary transformer aging [€] & 23.3 [23\%] & 65.5 [20\%] \\
Avoided secondary transformer aging [€] & 5.9 [6\%] & 29.1 [9\%] \\
Avoided cable aging [€] & 4.6 [4\%] & 42.2 [13\%] \\
Avoided MV losses [€] & 46.8 [45\%] & 134.2 [40\%] \\
Avoided LV losses [€] & 22.6 [22\%] & 64.5 [19\%] \\
Avoided voltage violations penalty$^{\dagger}$ & 720 & 4330 \\
\hline
Total avoided cost [€] & 103.2 & 335.5 \\
Market payments [€] & 92.4 & 258.4 \\
\textbf{Net welfare [€]} & 10.8 & 77.1 \\
\hline
\multicolumn{3}{p{0.98\linewidth}}{\footnotesize\emph{Note:} Activated-flexibility values are reported as Proposed / Per-node \cite{prat-network-aware} / 5-zone \cite{prat-network-aware}. The remaining economic metrics refer to the proposed method. $^{\dagger}$Soft voltage-penalty reduction is reported separately and is excluded from the net-welfare calculation.}
\end{tabular}
\end{table}

\begin{figure}
    \centering
    \includegraphics[width=1\linewidth]{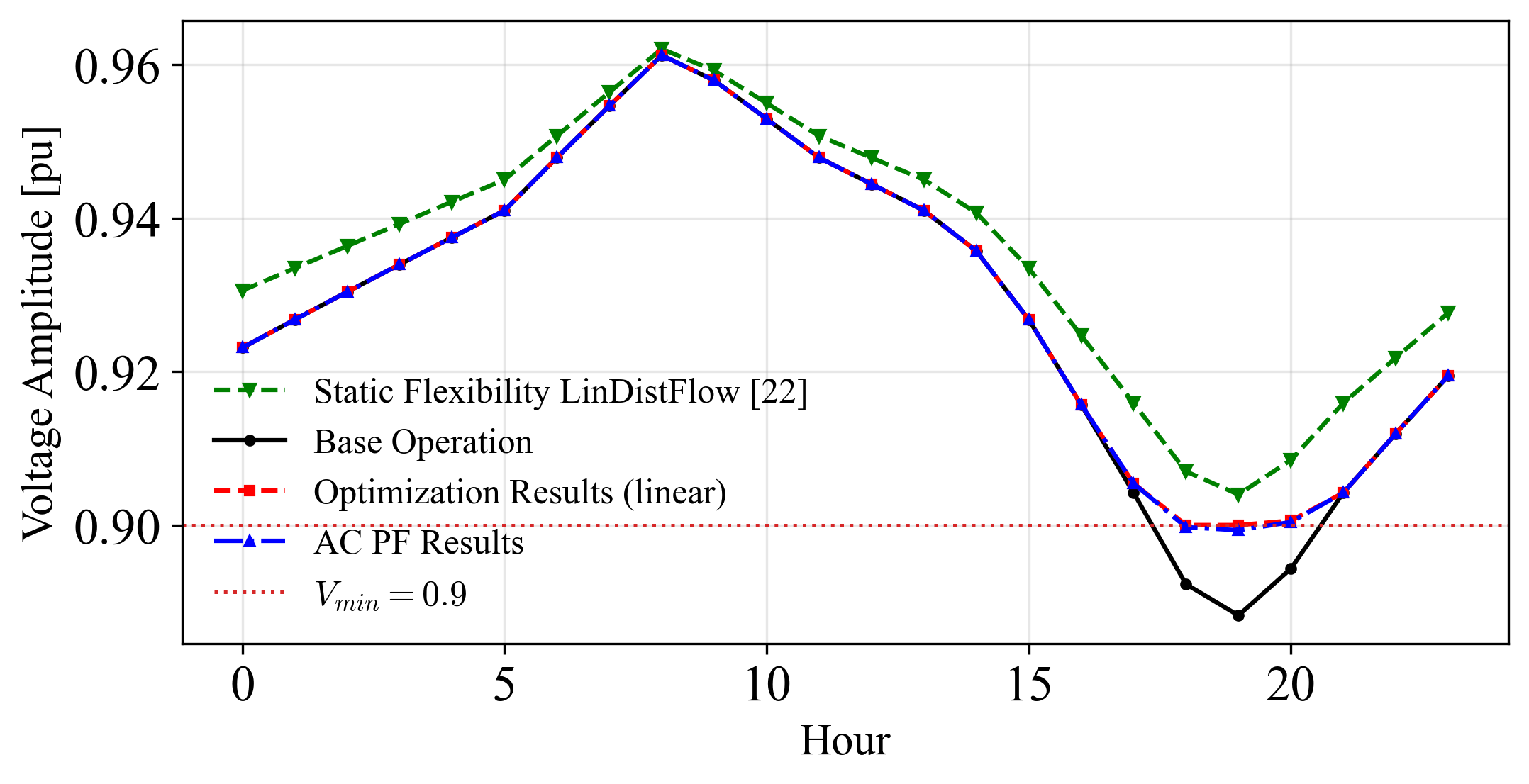}
    \caption{Comparison of the minimum voltage profile for the base-case scenario using the proposed methodology, the corresponding AC power-flow validation, the LinDistFlow-based flexibility quantification of \cite{prat-network-aware}, and the base operation without flexibility.}
    \label{fig:voltage_base_case_results}
\end{figure}

Finally, Fig.~\ref{fig:voltage_base_case_results} compares the minimum feeder voltage profile obtained with the proposed market-clearing framework, its corresponding full AC power-flow validation, the LinDistFlow-based flexibility quantification of \cite{prat-network-aware}, and the base operation without flexibility. The almost identical voltage trajectories of the proposed optimization and the AC validation confirm that the adopted sensitivity-based formulation accurately captures the network behavior for the resulting flexibility activation. \revision{In contrast, as discussed above, the LinDistFlow-based profile of \cite{prat-network-aware} does not correctly capture the true voltage depression and therefore gives safe operation, consistent with the inactive flexibility request.}

\subsection{Stress Case with 30\% EV Penetration}

To evaluate the proposed flexibility market's capability under future operating conditions, a stress scenario with 30\% EV penetration is investigated \revision{as defined in Section~\ref{sec:case_study}}. 
As summarized in Table~\ref{tab:market_results}, the additional EV demand increases the daily energy consumption of the examined network (14 secondary substations supplying 600 customers) by 2.7~MWh, while the system peak demand increases by approximately 600~kW during the evening peak at 20:00. Nevertheless, Fig.~3 of \cite{panagi2026zenodo} shows that, although the maximum loading increase at some secondary substations reaches nearly 50\%, the average increase during the critical evening hours remains close to only 10\%.

Despite this relatively modest increase in loading, the corresponding changes in DSO willingness to pay are significant. As illustrated in Fig.~\ref{fig:trafo2_split_lmp}, the peak loading of the most critical secondary transformer increases from approximately 0.9~p.u. under the base scenario to almost 1.05~p.u. in the stress case. Consequently, the proposed DLMP almost doubles, rising from approximately 65~€/MWh to more than 120~€/MWh. This behavior is primarily explained by the exponential dependence of transformer and cable thermal aging presented in \eqref{eq:faa_exact} and \eqref{eq:faa_cable_exact}, together with the quadratic dependence of electrical losses on loading. Therefore, relatively small increases in equipment loading can result in disproportionately larger marginal flexibility values.

\begin{figure}
    \centering
    \includegraphics[width=1\linewidth]{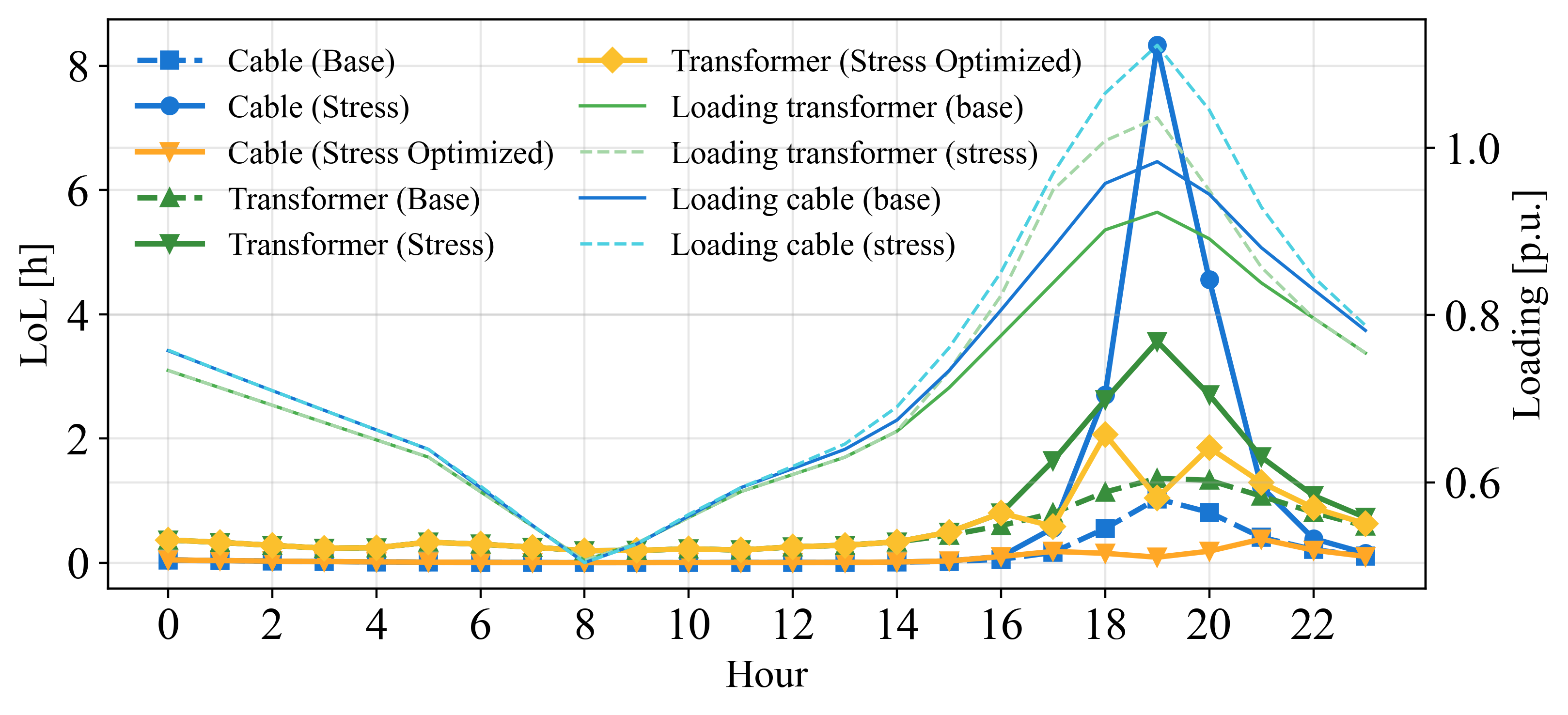}
    \caption{Hourly loss-of-life of a representative secondary transformer and MV cable under the base, stress, and optimized stress (flexibility activated) operating conditions.}
    \label{fig:lol_trafo2_idx0}
\end{figure}

The corresponding asset degradation is presented in Fig.~\ref{fig:lol_trafo2_idx0}. Under the base operating condition, both the transformer and the MV cable experience relatively low hourly loss-of-life values, indicating that operation is well within their thermal capabilities. Under the stress scenario, however, the accumulated daily transformer loss-of-life increases from 12.2~h to 19.4~h, while the cable loss-of-life increases from 3.6~h to 18.2~h. After applying the proposed flexibility market, these values are reduced to 13.7~h and 1.6~h, respectively. 

Additionally, the market-clearing results summarized in Table~\ref{tab:market_results} reveal an interesting shift in the composition of the DSO benefits. Compared to the base case, the contributions of secondary transformer and MV cable aging increase substantially, while the relative contributions of primary transformer aging and network loss reduction decrease. This observation demonstrates that the proposed willingness-to-pay formulation naturally adapts to prevailing network conditions, allocating greater economic value to components that become the dominant operational bottlenecks. Regarding voltage violations, almost full flexibility is enabled during congested time intervals to keep violations at the lowest achievable level.

\subsection{Sensitivity and Accuracy of Network Approximations}
\label{sec:network_approx}

\revision{The market-clearing formulation relies on first-order sensitivity approximations for voltage, line loading, and MV/LV losses around the anticipated baseline. This subsection examines two complementary aspects: (i)~the role of the voltage soft-penalty coefficient \(C\), and (ii)~the numerical discrepancy between the linearized network quantities used in the clearing and the corresponding full AC power-flow re-evaluation after activation, for both the base and stress cases.}

\subsubsection{Sensitivity of the Voltage Soft-Penalty Coefficient}
\label{sec:voltage_C_sensitivity}

\begin{figure}
    \centering
    \includegraphics[width=\linewidth]{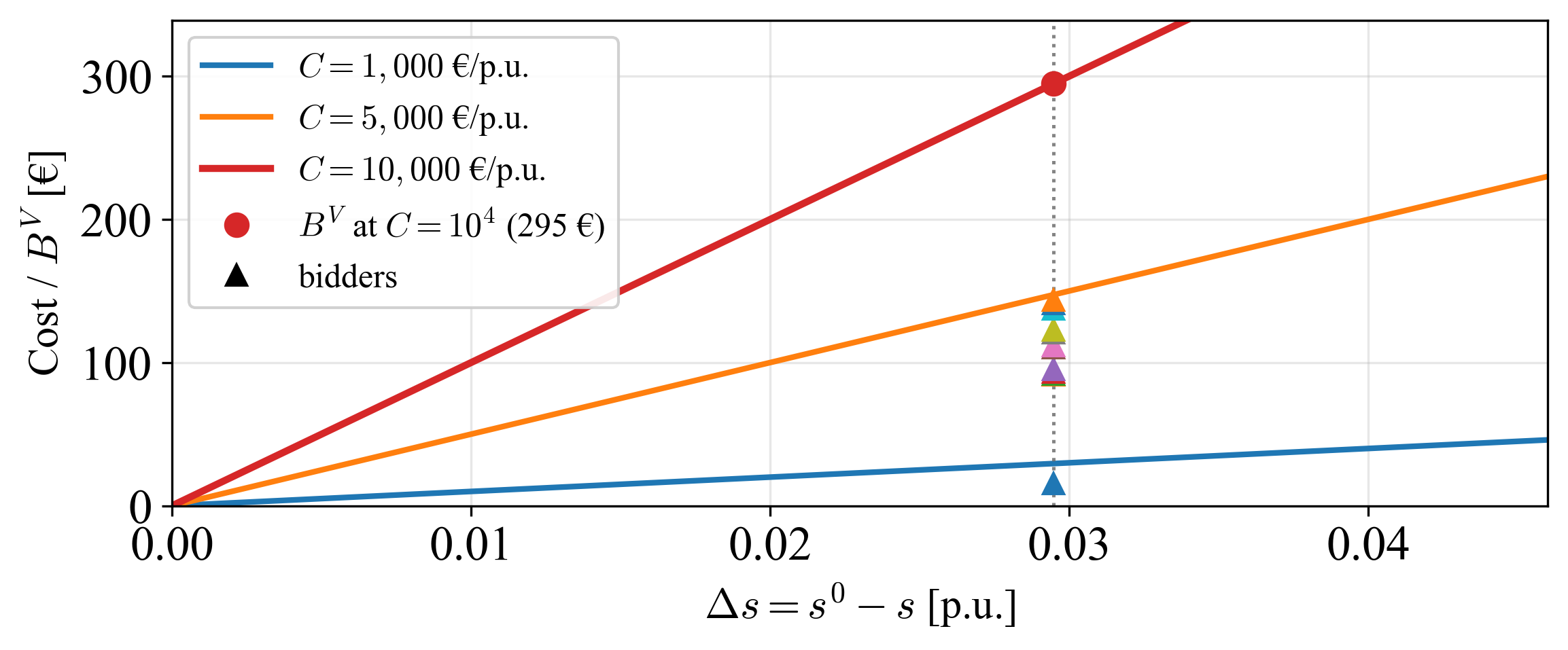}
    \caption{\revision{Illustrative sensitivity of the voltage soft-penalty term \(B^{\mathrm{V}}=C(s^{0}-s)\) at stress-case hour~19 (worst-bus undervoltage). Markers show FlexOffer costs of bidders for fully mitigating the same violation.}}
    \label{fig:voltage_C_sensitivity}
\end{figure}

\revision{To illustrate how the adopted soft-penalty coefficient \(C\) affects the voltage term relative to FlexOffer costs, Fig.~\ref{fig:voltage_C_sensitivity} considers the most severe undervoltage hour of the stress case (hour~19). For the worst bus, the baseline violation is \(s^{0}\approx 0.0295\)~p.u. The lines show \(B^{\mathrm{V}}=C\,(s^{0}-s)\) for \(C\in\{10^{3},5{\times}10^{3},10^{4}\}\)~\si{EUR/pu}, while the markers indicate the procurement cost that would be required from each bidder to fully remove that violation on its own. At the case-study value \(C=\SI{10000}{EUR/pu}\), the implied voltage benefit for full mitigation (\(\approx\SI{295}{EUR}\)) substantially exceeds all individual FlexOffer costs, so the soft penalty acts as a near-priority signal to relieve congestion whenever feasible offers exist. For smaller \(C\), \(B^{\mathrm{V}}\) falls into the range of the bid costs and the voltage term no longer dominates procurement. This supports treating \(C\) as an operational soft penalty in the present study and reporting the associated penalty reduction separately from the monetized net welfare.}

\subsubsection{Accuracy of the Sensitivity-Based Approximations}
\label{sec:approx_accuracy}

\revision{After market clearing, the accepted flexibility schedule is re-applied in a full AC power flow and compared with the linearized predictions used inside the optimization. Table~\ref{tab:approx_accuracy} reports this comparison for the base and stress cases through a small set of summary metrics. For transformers and MV lines, the maximum absolute error and the mean absolute error (MAE) of the per-unit loading $K_u$; for MV and LV losses, the absolute discrepancy of the cumulative daily loss-reduction energy; and for voltages, the maximum absolute bus-voltage error over all buses and hours. The LV loss row compares the first-order Taylor surrogate used in the clearing with the feeder-level quadratic model given in \eqref{eq:lv_losses}. Across all quantities, errors increase as we move from the base case to the stress case. This behavior is expected because the stress scenario involves larger flexibility requirements and, therefore, larger activations. As a result, the post-activation operating point moves farther from the baseline around which the first-order sensitivities are constructed, and the linearization residual naturally increases. Overall, while the discrepancies are generally limited relative to the cleared volumes, they can become substantial under strongly stressed activations, with some errors increasing by more than a factor of two. These results highlight the limitations of first-order network approximations when operating conditions deviate substantially from the initial linearization point.}

\begin{table}
\centering
\caption{\revision{Accuracy of the sensitivity-based network approximations against post-clearing AC power-flow validation.}}
\label{tab:approx_accuracy}
\begin{tabular}{lcc}
\hline
Quantity & Base case & Stress case \\
\hline
Transformer $K_u$: max $|\Delta K|$ [p.u.] & 0.0089 & 0.0140 \\
Transformer $K_u$: MAE [p.u.] & 0.0003 & 0.0008 \\
Line $K_u$: max $|\Delta K|$ [p.u.] & 0.0612 & 0.0998 \\
Line $K_u$: MAE [p.u.] & 0.0007 & 0.0016 \\
MV Losses $|E^{\mathrm{lin}}-E^{\mathrm{AC}}|$ [MWh] & 0.0183 & 0.0886 \\
LV Losses $|E^{\mathrm{lin}}-E^{k_s}|$ [MWh] & 0.0182 & 0.0586 \\
Max.\ $|V^{\mathrm{lin}}-V^{\mathrm{AC}}|$ [p.u.] & 0.0024 & 0.0049 \\
\hline
\end{tabular}
\end{table}

\subsection{Comparison and Discussion}

The obtained results highlight several findings. First, the proposed piecewise-linear aging formulation allows aging models to be incorporated while retaining a linear, convex market-clearing formulation and achieving accuracy nearly identical to that of the exact nonlinear transformer and cable thermal models. This enables the explicit incorporation of asset degradation into the market-clearing process without sacrificing computational tractability. Moreover, the base-case results demonstrate that even under relatively uncongested operating conditions, flexibility can still provide measurable economic benefits by reducing transformer and cable aging together with network losses, creating value for both the DSO and the flexibility service providers.

\begin{figure}
    \centering
    \includegraphics[width=1\linewidth]{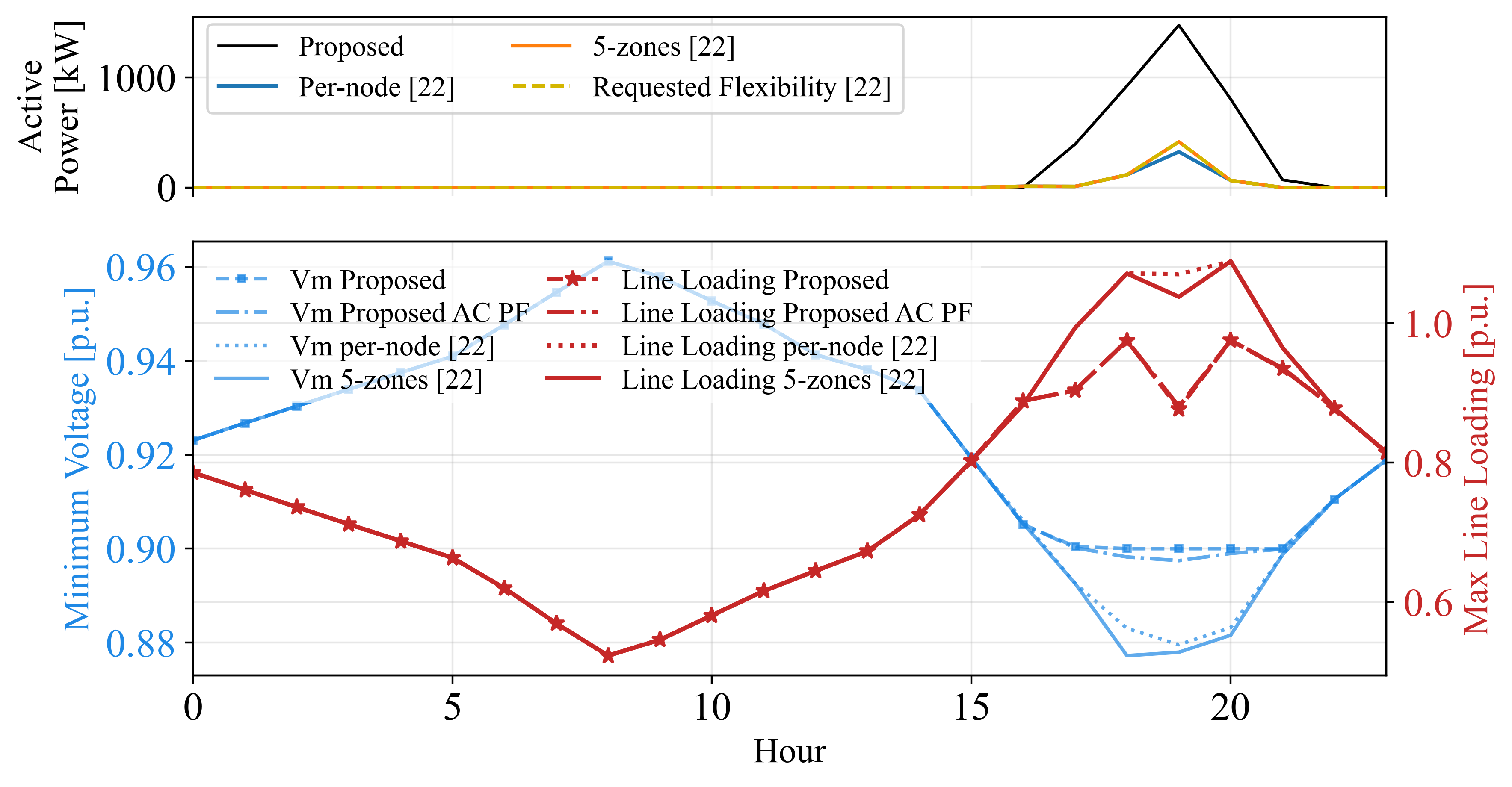}
    \caption{Comparison of the proposed methodology with the benchmark approaches of \cite{prat-network-aware} in terms of activated flexibility (top), minimum feeder voltage (bottom, left axis), and maximum line loading (bottom, right axis) under the stress-case operating conditions.}
    \label{fig:comparison}
\end{figure}

More importantly, the proposed framework fundamentally differs from existing flexibility-request methodologies by replacing predefined static flexibility requests with a dynamic DSO willingness-to-pay embedded directly into the market-clearing objective. As illustrated in Fig.~\ref{fig:comparison}, the benchmark methodology of \cite{prat-network-aware} requests flexibility only to alleviate anticipated congestion, resulting in significantly lower market participation. In contrast, the proposed approach activates flexibility whenever the overall network benefit exceeds the procurement cost, thereby exploiting additional economic opportunities beyond congestion management. Consequently, market liquidity is substantially increased while preserving the transparency and privacy-preserving characteristics of the market architecture. Furthermore, the benchmark-per-node formulation yields the most localized flexibility requests and limits the number of eligible participants. Expanding these requests into five flexibility zones increases market participation but leads to poorer technical performance, with remaining voltage and line-loading violations after AC power-flow validation. Conversely, the proposed methodology maintains network operation close to the desired operating limits.

\revision{While these results demonstrate the improved technical performance of the proposed clearing mechanism, the network representation's thermal loading warrants further investigation. In the present clearing formulation, transformer and cable loading enter only through the monetized aging terms; no explicit line-loading or temperature security constraints are imposed. An explicit loading penalty could, in principle, be introduced in a sensitivity-based manner similar to voltage operational terms. However, we omit such a term here because the aging model already provides a strong economic incentive when assets operate close to overload, as confirmed by the base-versus-stress comparison. The aging benefit increases sharply under higher loading, so the willingness to pay adapts without an additional arbitrary loading coefficient.}

\revision{An additional important aspect is the local character of LFMs, which raises the possibility of market power and strategic bidding when only a few providers are electrically well placed to relieve a constraint. On the algorithmic side, as discussed above, the proposed dynamic clearing does not pre-select a bus or zone and can therefore draw on offers from all participating locations. Favorable sites may still exist, yet the broader pool of eligible offers can increase liquidity and competitive pressure, making it less likely that a single location will remain the only viable option. Complementary market-design mechanisms, such as price caps, profit or mark-up limits informed by estimated aggregator benefits, and regulatory oversight of abusive bidding, can further limit locational exploitation beyond the clearing algorithm itself.}

\section{Conclusions}
\label{sec:conclusion}

This paper presents a dynamic flexibility-request methodology for LFMs, in which the DSO's willingness-to-pay is determined directly within the market-clearing process rather than relying on predefined flexibility requests. To support this formulation, exact convex piecewise-linear reformulations of the transformer and cable thermal aging models are developed, allowing asset degradation to be explicitly considered together with network losses and voltage congestion while preserving a linear market-clearing formulation.
The results demonstrate that, unlike conventional congestion-driven flexibility requests, the proposed methodology successfully identifies economically beneficial flexibility even under lightly congested operating conditions, thereby significantly increasing market liquidity and improving overall social welfare. Comparisons with a recent state-of-the-art methodology further show improved flexibility utilization and enhanced congestion management.

Future work will focus on extending the proposed framework through an iterative market-clearing process to further reduce the small AC feasibility errors introduced by sensitivity-based linearization under large flexibility activations. In addition, more detailed flexibility-provider models, including internal operational constraints and realistic resource aggregation, will be incorporated into the market-clearing problem. Finally, future research will investigate the design and bidding strategies for flexibility offers, enabling a more comprehensive representation of both DSO and FSP decision-making.
\section*{Acknowledgment}
This project has received funding from the European Union’s Framework Programme for Research and Innovation Horizon Europe under the Marie Skłodowska-Curie Actions Doctoral Networks, Grant Agreement No. 101120278.

\bibliographystyle{IEEEtran}
\bibliography{biblio} 
{\appendices
\section{Proof of the Exactness of the Proposed Epigraph-Based PWL Reformulation}
\label{app:conv_monot_proof}

Consider the thermal-rise maps $f_1(K)$ and $f_2(K)$ in~\eqref{eq:TO_ultimate}--\eqref{eq:H_ultimate} and the aging acceleration factor $F_{\mathrm{AA}}(\theta^H)$ in~\eqref{eq:faa_exact}. Throughout this appendix we assume $40^\circ\mathrm{C}\le\theta^H\le140^\circ\mathrm{C}$, $R>0$, $n,m\ge\tfrac12$, and $K\ge0$, which cover the transformer parameters used in this work~\cite{6166928}.

\subsection{Convexity and Exactness of the Epigraph Reformulation}
Under these assumptions the second derivatives satisfy $\frac{d^2f_1}{dK^2}\ge0$, $\frac{d^2f_2}{dK^2}\ge0$, and $\frac{d^2F_{\mathrm{AA}}}{d(\theta^H)^2}>0$; hence $f_1$, $f_2$, and $F_{\mathrm{AA}}$ are convex on the operating domain and admit convex PWL approximations.

The thermal dynamics are strictly monotone in the ultimate rises: $\partial\Delta\theta^{TO}/\partial\Delta\theta^{TO,U}=\alpha^{TO}=1-e^{-\Delta t/\tau^{TO}}\in(0,1)$
and $\partial\Delta\theta^{H}/\partial\Delta\theta^{H,U}=\alpha^{H}=1-e^{-\Delta t/\tau^{H}}\in(0,1)$
for $\Delta t,\tau^{TO},\tau^{H}>0$. Moreover, $\theta^H=\theta^A+\Delta\theta^{TO}+\Delta\theta^{H}$ implies $\partial\theta^H/\partial\Delta\theta^{TO}=\partial\theta^H/\partial\Delta\theta^{H}=1>0$, while $dF_{\mathrm{AA}}/d\theta^H>0$. Consequently, $F_{\mathrm{AA}}$ is strictly increasing along the thermal chain, and the monotonicity relations \eqref{eq:monotonic_TO} and \eqref{eq:monotonic_H} hold.
\begin{align}
\label{eq:monotonic_TO}
\Delta\theta^{TO,U} &\uparrow \Longrightarrow \Delta\theta^{TO} \uparrow \Longrightarrow \theta^H \uparrow \Longrightarrow F_{\mathrm{AA}} \uparrow, \\
\label{eq:monotonic_H}
\Delta\theta^{H,U} &\uparrow \Longrightarrow \Delta\theta^{H} \uparrow \Longrightarrow \theta^H \uparrow \Longrightarrow F_{\mathrm{AA}} \uparrow
\end{align}

Therefore, due to the convexity, if the auxiliary FAA cost is minimized in the objective, any inactive PWL epigraph inequality for $f_1$ or $f_2$ would allow a feasible reduction of the corresponding auxiliary variable that strictly improves the objective, contradicting optimality (Section~\ref{sec:epigraph_reformulation}). Hence, at least one inequality of every epigraph set is active at the optimum, and the reformulation is exact up to the PWL breakpoint error.

\subsection{Exactness under the Complete Market-Clearing Objective}

The established exactness of the thermal-aging epigraph formulation also holds for the full market-clearing objective, despite the simultaneous impact of flexibility decisions on losses, voltages, and asset loading. Let $\mathbf{x}$ be the market-clearing variables (including $\Delta P$), and $\mathbf{y}$ the auxiliary thermal-aging variables. The full objective is given in \eqref{eq:complete_market_exactness},
\begin{equation}
    \label{eq:complete_market_exactness}
    \max_{\mathbf{x},\mathbf{y}}
    \quad
    C_{\mathrm{age}}^{0}
    -
    C_{\mathrm{age}}^{\mathrm{opt}}(\mathbf{x},\mathbf{y})
    +
    B_{\mathrm{other}}(\mathbf{x})
    -
    C_{\mathrm{FSP}}(\mathbf{x})
    \end{equation}
where $B_{\mathrm{other}}$ includes the losses and voltage-related benefits. 

Consider an epigraph variable $y_r$ associated with a convex PWL approximation of a thermal function. Its feasible region is defined by $y_r\ge a_{r,k} z_r+b_{r,k} \forall k\in\mathcal{K}_r$. Suppose, by contradiction, that at an optimal solution $(\mathbf{x}^{\star},\mathbf{y}^{\star})$, $y_r^{\star}
> \max_{k\in\mathcal{K}_r} \left\{a_{r,k} z_r^{\star}+b_{r,k}\right\}$. In this case, $y_r$ can be reduced to $\widehat{y}_r = \max_{k\in\mathcal{K}_r} \left\{ a_{r,k} z_r^{\star} + b_{r,k} \right\}$, without altering $\mathbf{x}^{\star}$ or violating any epigraph constraint. Therefore, the losses, voltage benefits, and FSP payments remain unchanged, whereas the optimized aging cost decreases. This implies $C_{\mathrm{age}}^{\mathrm{opt}}(\widehat{\mathbf{y}}) < C_{\mathrm{age}}^{\mathrm{opt}}(\mathbf{y}^{\star})$, which contradicts the optimality of $(\mathbf{x}^{\star},\mathbf{y}^{\star})$. Thus, for every thermal epigraph variable, $y_r^{\star} = \max_{k\in\mathcal{K}_r} \left\{ a_{r,k} z_r^{\star} + b_{r,k} \right\}$, meaning at least one PWL segment is active at the optimum. Consequently, the complete market-clearing formulation introduces no epigraph relaxation gap.

\hfill$\blacksquare$
}

\end{document}